\documentclass[byrevtex,showpacs,showkeys,aps,twocolumn]{revtex4}%
\pdfoutput=1
\usepackage{amsfonts}
\usepackage{amsmath}
\usepackage{amssymb}
\usepackage{graphicx}%
\providecommand{\U}[1]{\protect\rule{.1in}{.1in}}
\newtheorem{theorem}{Theorem}

\newtheorem{corollary}[theorem]{Corollary}

\newenvironment{proof}[1][Proof]{\noindent\textbf{#1.} }{\ \rule{0.5em}{0.5em}}
\begin{document}
\preprint{ }
\title[ ]{Quantum Shift Register Circuits}
\author{Mark M. Wilde}
\affiliation{Electronic Systems Division, Science Applications International Corporation,
4001 North Fairfax Drive, Arlington, Virginia, USA\ 22203}

\begin{abstract}
A quantum shift register circuit acts on a set of input qubits and memory
qubits, outputs a set of output qubits and updated memory qubits, and feeds
the memory back into the device for the next cycle (similar to the operation
of a classical shift register). Such a device finds application as an encoding
and decoding circuit for a particular type of quantum error-correcting code,
called a quantum convolutional code. Building on the Ollivier-Tillich and
Grassl-R\"{o}tteler encoding algorithms for quantum convolutional codes, I
present a method to determine a quantum shift register encoding circuit for a
quantum convolutional code. I also determine a formula for the amount of
memory that a CSS\ quantum convolutional code requires. I then detail
primitive quantum shift register circuits that realize all of the finite- and
infinite-depth transformations in the shift-invariant Clifford group (the
class of transformations important for encoding and decoding quantum
convolutional codes). The memory formula for a CSS\ quantum convolutional code
then immediately leads to a formula for the memory required by a
CSS\ entanglement-assisted quantum convolutional code.

\end{abstract}
\volumeyear{2008}
\volumenumber{ }
\issuenumber{ }
\eid{ }
\date{\today}
\startpage{1}
\endpage{10}
\maketitle

\section{Introduction}

With the advent of quantum computing and quantum communication, it becomes
increasingly important to develop ways for protecting quantum information
against the adversarial effects of noise \cite{qecbook}. Researchers have
developed many theoretical techniques for the protection of quantum
information
\cite{PhysRevLett.77.793,PhysRevA.54.1098,thesis97gottesman,PhysRevLett.78.405,ieee1998calderbank,PhysRevLett.79.3306,mpl1997zanardi,PhysRevLett.81.2594,kribs:180501,qic2006kribs,poulin:230504,arxiv2007brun}%
\ since Shor's original contribution to the theory of quantum error correction
\cite{PhysRevA.52.R2493}.

Quantum convolutional coding is a technique for protecting a stream of quantum
information
\cite{PhysRevLett.91.177902,arxiv2004olliv,isit2006grassl,ieee2006grassl,ieee2007grassl,isit2005forney,ieee2007forney,cwit2007aly,arx2007aly,arx2007wildeCED,arx2007wildeEAQCC,arx2008wildeUQCC,arx2008wildeGEAQCC,pra2009wilde}
and is perhaps more valuable for quantum communication than it is for quantum
computation (though see the tail-biting technique in
Ref.~\cite{ieee2007forney}). Quantum convolutional codes bear similarities to
classical convolutional codes \cite{book1999conv,mct2008book}. The encoding
circuit for a quantum convolutional code consists of a single unitary
repeatedly applied to the quantum data stream \cite{PhysRevLett.91.177902}.
Decoding a quantum convolutional code consists of applying a syndrome-based
version of the Viterbi decoding algorithm
\cite{itit1967viterbi,ieee2007forney,PhysRevLett.91.177902}.

The encoding circuit for a classical convolutional code has a particularly
simple form. Given a mathematical description of a classical convolutional
code, one can easily write down a shift register implementation for the
encoding circuit \cite{book1999conv}. For this reason among others, deep space
missions such as \textit{Voyager} and \textit{Pioneer }used classical
convolutional codes to protect classical information \cite{ieee2007pollara}.

A natural question is whether there exists such a simple mapping from the
mathematical description of a quantum convolutional code to a quantum shift
register implementation. Many researchers have investigated the mathematical
constructions of quantum convolutional codes, but few
\cite{arxiv2004olliv,isit2006grassl,ieee2006grassl,ieee2007grassl}\ have
attempted to develop encoding circuits for them. The Ollivier-Tillich quantum
convolutional encoding algorithm \cite{arxiv2004olliv} is similar to
Gottesman's technique \cite{thesis97gottesman}\ for encoding a quantum block
code. The Grassl-R\"{o}tteler encoding algorithm
\cite{isit2006grassl,ieee2006grassl,ieee2007grassl} encodes a quantum
convolutional code with a sequence of elementary encoding operations. Each of
these elementary encoding operations has a mathematical representation as a
polynomial matrix, and each elementary encoding operation builds up the
mathematical representation of the quantum convolutional code.

The Ollivier-Tillich and Grassl-R\"{o}tteler encoding algorithms leave a
practical question unanswered. They both do not determine how much memory a
given encoding circuit requires, and in the Grassl-R\"{o}tteler algorithm, it
is not even explicitly clear how the encoding circuit obeys a convolutional
structure (it obeys a periodic structure, but the convolutional structure
demands that the encoding circuit consist of the same single unitary applied
repeatedly on the quantum data stream).

In this paper, I develop the theory of quantum shift register circuits, using
tools familiar from linear system theory \cite{kalaith1980book}\ and classical
convolutional codes \cite{book1999conv}. I explicitly show how to connect
quantum shift register circuits together so that they encode a quantum
convolutional code. I develop a general technique for reducing the amount of
memory that the quantum shift register encoding circuit requires.
Theorem~\ref{thm:memory-CSS}\ of this paper answers the above question
concerning memory use in a CSS\ quantum convolutional code---it determines the
amount of memory that a given CSS\ quantum convolutional code requires, as a
function of the mathematical representation of the code. I also show how to
implement any elementary operation from the shift-invariant Clifford group
\cite{isit2006grassl,arx2007wildeEAQCC}\ with a quantum shift register circuit.

These quantum shift register circuits might be of interest to experimentalists
wishing to implement a quantum error-correcting code that has a simple
encoding circuit, but unlike a quantum block code, has a memory structure.
Classical convolutional codes were most useful in the early days of computing
and communication because they have a higher performance/complexity trade-off
than a block code that encodes the same number of information qubits
\cite{ieee2007forney}. At the current stage of development, experimentalists
have the ability to perform few-qubit interactions, and it might be useful to
exploit these few-qubit interactions on a quantum data stream, rather than on
a single block of qubits.

Other authors have suggested the idea of a quantum shift register
\cite{prsa2000grassl,arx2001QSR}, but it is not clear how we can apply the
ideas in these papers to the encoding of a quantum convolutional code.
Additionally, another set of authors labeled their work as a \textquotedblleft
quantum shift register\textquotedblright\ \cite{ieee2001QSR},\ but this
quantum shift register is not useful for protecting quantum information (nor
is it even useful for coherent quantum operations).

The closest work to this one is the discussion in Section~IIB\ of
Ref.~\cite{arx2007poulin}. Though, Poulin \textit{et al.} did not develop the
quantum shift register idea in much detail because their focus was to develop
the theory of decoding quantum turbo codes. The discussion in
Ref.~\cite{arx2007poulin} is one of the inspirations for this work (as well as
the initial work of Ollivier and Tillich
\cite{PhysRevLett.91.177902,arxiv2004olliv}), and this paper is an extension
of that discussion.

The most natural implementation of a quantum shift register circuit may be in
a spin chain \cite{PhysRevLett.91.207901,B07}. Such an implementation requires
a repetition of acting with the encoding unitary at the sender's register and
allowing the Hamiltonian of the spin chain to shift the qubits by a certain
amount. Further investigation is necessary to determine if this scheme would
be feasible. Another natural implementation of a quantum shift register
circuit is with linear optical circuits \cite{Knill:2001:46}. One can
implement the feedback necessary for this circuit by redirecting light beams
with mirrors. The difficulty with this approach is that controlled-unitary
encoding is probabilistic.

I structure this work as follows. The next section begins with examples that
illustrate the operation of a quantum shift register circuit. I then present a
simple example of a quantum shift register circuit that encodes a CSS\ quantum
convolutional code. This example demonstrates the main ideas for constructing
quantum shift register encoding circuits. First, build a quantum shift
register circuit for each elementary encoding operation in the
Grassl-R\"{o}tteler encoding algorithm. Then, connect the outputs of the first
quantum shift register circuit to the inputs of the next one and so on for all
of the elementary quantum shift register circuits. Finally, simplify the
device by determining how to \textquotedblleft commute gates through the
memory\textquotedblright\ of the larger quantum shift register circuit
(discussed in more detail later). This last step allows us to reduce the
amount of memory that the quantum shift register circuit requires.
Section~\ref{sec:finite-depth-CNOTs} follows this example by developing two
types of finite-depth controlled-NOT (CNOT)\ quantum shift register circuits
(I explain the definition of \textquotedblleft finite-depth\textquotedblright%
\ later on). Section~\ref{sec:memory-comp}\ then states and proves
Theorem~\ref{thm:memory-CSS}---this theorem gives a formula to determine the
amount of memory that a given CSS\ quantum convolutional code requires. I then
develop the theory of quantum shift register circuits with controlled-phase
gates and follow by giving the encoding circuit for the Forney-Grassl-Guha
code ~\cite{ieee2007forney}. Grassl and R\"{o}tteler stated that the encoding
circuit for this code requires two frames of memory qubits
\cite{isit2006grassl}, but I instead find with this paper's technique that the
minimum amount it requires is five frames. Section~\ref{sec:infinite-depth}
then develops quantum shift register circuits for infinite-depth operations,
which are important for the encoding of Type II CSS\ entanglement-assisted
quantum convolutional codes \cite{arx2007wildeEAQCC}.
Theorem~\ref{thm:memory-CSS}\ also determines the amount of memory required by
these codes. I then conclude with some observations and open questions.

\section{Examples of Quantum Shift Register Circuits}

Let us begin with a simple example to show how we can build up an arbitrary
finite-depth CNOT operation. Consider the full set of Pauli operators on two
qubits \cite{qecbook,book2000mikeandike}:%
\[%
\begin{array}
[c]{cc}%
Z & I,\\
I & Z,\\
X & I,\\
I & X.
\end{array}
\]
We can form a symplectic representation of the full set of Pauli operators for
two qubits with the following matrix \cite{qecbook,book2000mikeandike}:%
\[
\left[  \left.
\begin{array}
[c]{cc}%
1 & 0\\
0 & 1\\
0 & 0\\
0 & 0
\end{array}
\right\vert
\begin{array}
[c]{cc}%
0 & 0\\
0 & 0\\
1 & 0\\
0 & 1
\end{array}
\right]  ,
\]
where the entries to the left of the vertical bar correspond to the $Z$
operators and the entries to the right of the vertical bar correspond to the
$X$ operators. Suppose that we perform a CNOT\ gate from the first qubit to
the second qubit conditional on a bit $f_{0}$. We perform the gate if
$f_{0}=1$ and do not perform it otherwise. The above Pauli operators transform
as follows:%
\[
\left[  \left.
\begin{array}
[c]{cc}%
1 & 0\\
f_{0} & 1\\
0 & 0\\
0 & 0
\end{array}
\right\vert
\begin{array}
[c]{cc}%
0 & 0\\
0 & 0\\
1 & f_{0}\\
0 & 1
\end{array}
\right]  .
\]%
%TCIMACRO{\FRAME{ftbpFU}{2.0358in}{0.5812in}{0pt}{\Qcb{The above figure depicts
%a simple CNOT\ transformation conditional on the bit $f_{0}$. The circuit does
%not apply the gate if $f_{0}=0$ and applies it if $f_{0}=1$.}}%
%{\Qlb{fig:simple-CNOT}}{simple-cnot.pdf}%
%width 2.0358in;  height 0.5812in;  depth 0pt;  original-width 3.8536in;
%original-height 1.0603in;  cropleft "0";  croptop "1";  cropright "1";
%cropbottom "0";  filename 'figures/simple-CNOT.pdf';file-properties "XNPEU";}%
%}}%
%BeginExpansion
\begin{figure}
[ptb]
\begin{center}
\includegraphics[
natheight=1.060300in,
natwidth=3.853600in,
height=0.5812in,
width=2.0358in
]%
{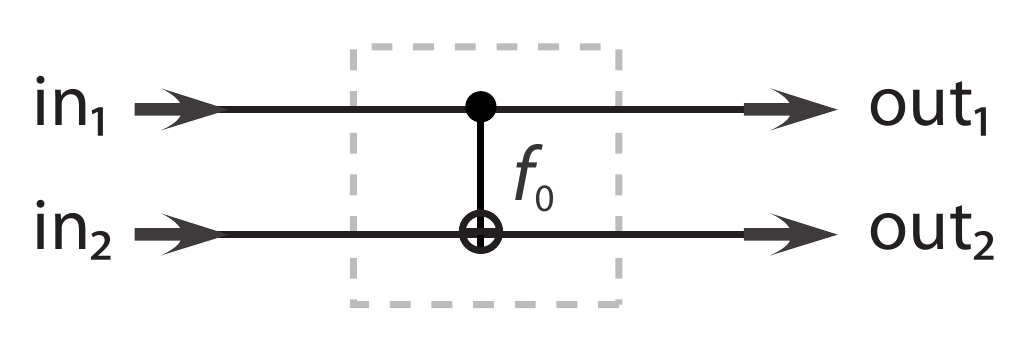}%
\caption{The above figure depicts a simple CNOT\ transformation conditional on
the bit $f_{0}$. The circuit does not apply the gate if $f_{0}=0$ and applies
it if $f_{0}=1$.}%
\label{fig:simple-CNOT}%
\end{center}
\end{figure}
%EndExpansion
Figure~\ref{fig:simple-CNOT}\ depicts the \textquotedblleft quantum shift
register circuit\textquotedblright\ that implements this transformation (this
device is not really a quantum shift register circuit because it does not
exploit a set of memory qubits).

\begin{figure}
[ptb]
\begin{center}
\includegraphics[
natheight=2.559800in,
natwidth=4.726200in,
height=1.1156in,
width=2.0358in
]%
{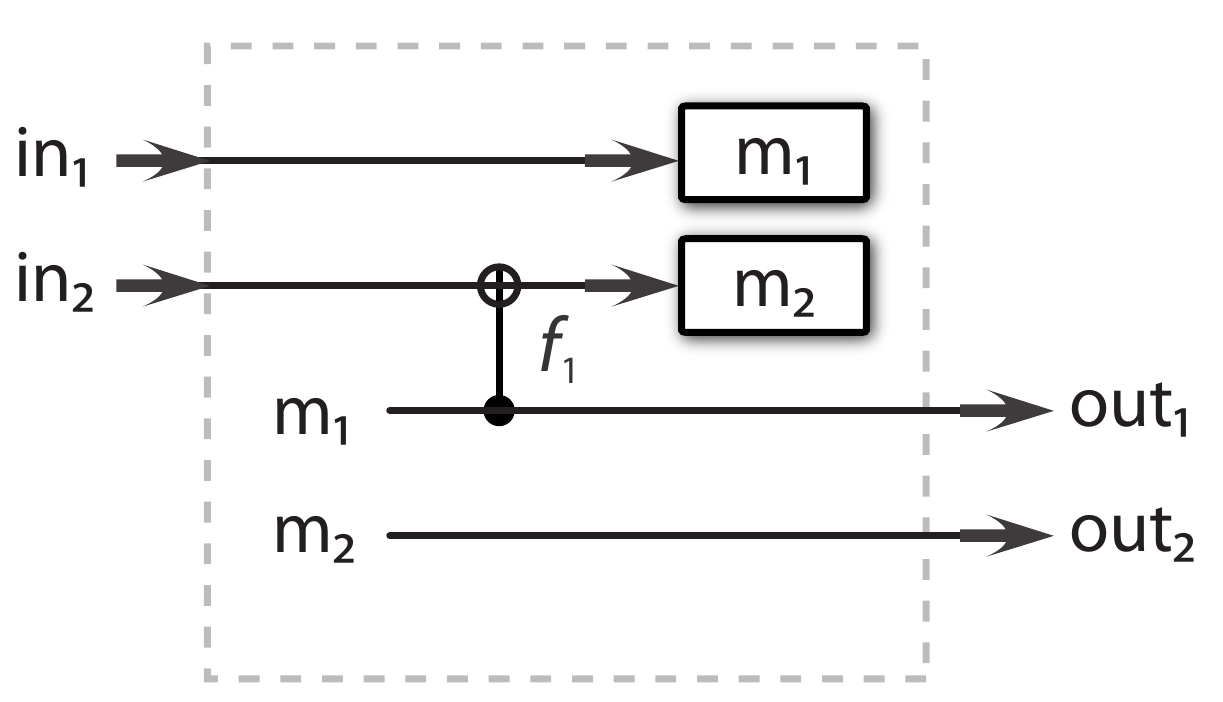}%
\caption{A quantum shift register device that incorporates one frame of memory
qubits.}%
\label{fig:one-delay-CNOT}%
\end{center}
\end{figure}

Let us incorporate one frame of memory qubits so that the circuit really now
becomes a quantum shift register circuit. Consider the circuit in
Figure~\ref{fig:one-delay-CNOT}. The first two qubits are fed into the device
and the second one is the target of a CNOT\ gate from a future frame of qubits
(conditional on the bit $f_{1}$). The two qubits are then stored as two memory
qubits (swapped out with what was previously there). On the next cycle, the
two qubits are fed out and the first qubit that was previously in memory acts
on the second qubit in a frame that is in the past with respect to itself. We
would expect the $X$ variable of the first outgoing qubit to propagate one
frame into the past with respect to itself and the $Z$ variable of the second
incoming qubit to propagate one frame into the future with respect to itself.
We make this idea more clear in the below analysis.%
%TCIMACRO{\FRAME{ftbpFU}{2.0358in}{1.1156in}{0pt}{\Qcb{A quantum shift register
%device that incorporates one frame of memory qubits.}}%
%{\Qlb{fig:one-delay-CNOT}}{one-delay-cnot.pdf}%
%width 2.0358in;  height 1.1156in;  depth 0pt;  original-width 4.7262in;
%original-height 2.5598in;  cropleft "0";  croptop "1";  cropright "1";
%filename 'figures/one-delay-CNOT.pdf';file-properties "XNPEU";}}}%
%BeginExpansion
%EndExpansion

We can analyze this situation with a set of recursive equations. Let
$x_{1}\left[  n\right]  $ denote the bit representation of the $X$ Pauli
operator for the first incoming qubit at time $n$ and let $z_{1}\left[
n\right]  $ denote the bit representation of the $Z$ Pauli operator for the
first incoming qubit at time $n$. Let $x_{2}\left[  n\right]  $ and
$z_{2}\left[  n\right]  $ denote similar quantities for the second incoming
qubit at time $n$. Let $m_{1}^{x}\left[  n\right]  $ denote the bit
representation of the $X$ Pauli operator acting on the first memory qubit at
time $n$ and let $m_{1}^{z}\left[  n\right]  $ denote the bit representation
of the $Z$ Pauli operator acting on the first memory qubit at time $n$. Let
$m_{2}^{x}\left[  n\right]  $ and $m_{2}^{z}\left[  n\right]  $ denote similar
quantities for the second memory qubit. In the symplectic bit vector notation,
we denote the \textquotedblleft Z\textquotedblright\ part of the Pauli
operators acting on these four qubits at time $n$ as%
\[
\mathbf{z}\left[  n\right]  \equiv\left[
\begin{array}
[c]{cccc}%
z_{1}\left[  n\right]  & z_{2}\left[  n\right]  & m_{1}^{z}\left[  n-1\right]
& m_{2}^{z}\left[  n-1\right]
\end{array}
\right]  ,
\]
and the \textquotedblleft X\textquotedblright\ part by%
\[
\mathbf{x}\left[  n\right]  \equiv\left[
\begin{array}
[c]{cccc}%
x_{1}\left[  n\right]  & x_{2}\left[  n\right]  & m_{1}^{x}\left[  n-1\right]
& m_{2}^{x}\left[  n-1\right]
\end{array}
\right]  .
\]
The symplectic vector for the inputs is then%
\begin{equation}
\left[  \left.
\begin{array}
[c]{c}%
\mathbf{z}\left[  n\right]
\end{array}
\right\vert
\begin{array}
[c]{c}%
\mathbf{x}\left[  n\right]
\end{array}
\right]  . \label{eq:symplectic-vector}%
\end{equation}
I prefer this bit notation of Poulin \textit{et al}. \cite{arx2007poulin}
because it is more flexible for quantum shift register circuits. It allows us
to capture the evolution of an arbitrary tensor product of Pauli operators
acting on these four qubits at time $n$.

At time $n$, the two incoming qubits and the \textit{previous} memory qubits
from time $n-1$ are fed into the quantum shift register device and the
CNOT\ gate acts on them. The notation in Figure~\ref{fig:one-delay-CNOT}
indicates that there is an implicit swap at the end of the operation. The
incoming qubits get fed into the memory, and the previous memory qubits get
fed out as output. Let $x_{1}^{\prime}\left[  n\right]  $, $z_{1}^{\prime
}\left[  n\right]  $, $x_{2}^{\prime}\left[  n\right]  $, and $z_{2}^{\prime
}\left[  n\right]  $ denote the respective output variables. The symplectic
transformation for the CNOT\ gate is%
\[
\left[  \left.
\begin{array}
[c]{cccc}%
1 & 0 & 0 & 0\\
0 & 1 & f_{1} & 0\\
0 & 0 & 1 & 0\\
0 & 0 & 0 & 1\\
0 & 0 & 0 & 0\\
0 & 0 & 0 & 0\\
0 & 0 & 0 & 0\\
0 & 0 & 0 & 0
\end{array}
\right\vert
\begin{array}
[c]{cccc}%
0 & 0 & 0 & 0\\
0 & 0 & 0 & 0\\
0 & 0 & 0 & 0\\
0 & 0 & 0 & 0\\
1 & 0 & 0 & 0\\
0 & 1 & 0 & 0\\
0 & f_{1} & 1 & 0\\
0 & 0 & 0 & 1
\end{array}
\right]  .
\]
The above matrix postmultiplies the vector in (\ref{eq:symplectic-vector}) to
give the following output vector. We denote the \textquotedblleft
Z\textquotedblright\ part of the output Pauli operators acting on these four
qubits at time $n$ as%
\[
\mathbf{z}^{\prime}\left[  n\right]  \equiv\left[
\begin{array}
[c]{cccc}%
m_{1}^{z}\left[  n\right]  & m_{2}^{z}\left[  n\right]  & z_{1}^{\prime
}\left[  n\right]  & z_{2}^{\prime}\left[  n\right]
\end{array}
\right]  ,
\]
and the \textquotedblleft X\textquotedblright\ part by%
\[
\mathbf{x}^{\prime}\left[  n\right]  \equiv\left[
\begin{array}
[c]{cccc}%
m_{1}^{x}\left[  n\right]  & m_{2}^{x}\left[  n\right]  & x_{1}^{\prime
}\left[  n\right]  & x_{2}^{\prime}\left[  n\right]
\end{array}
\right]  ,
\]
with the change of locations corresponding to the implicit swap. The
symplectic vector for the outputs is then%
\begin{equation}
\left[  \left.
\begin{array}
[c]{c}%
\mathbf{z}^{\prime}\left[  n\right]
\end{array}
\right\vert
\begin{array}
[c]{c}%
\mathbf{x}^{\prime}\left[  n\right]
\end{array}
\right]  .
\end{equation}
It is simpler to describe the above transformation as a set of recursive
\textquotedblleft update\textquotedblright\ equations:%
\begin{align*}
x_{1}^{\prime}\left[  n\right]   &  =m_{1}^{x}\left[  n-1\right]  ,\\
z_{1}^{\prime}\left[  n\right]   &  =m_{1}^{z}\left[  n-1\right]  +f_{1}%
z_{2}\left[  n\right]  ,\\
x_{2}^{\prime}\left[  n\right]   &  =m_{2}^{x}\left[  n-1\right]  ,\\
z_{2}^{\prime}\left[  n\right]   &  =m_{2}^{z}\left[  n-1\right]  ,\\
m_{1}^{x}\left[  n\right]   &  =x_{1}\left[  n\right]  ,\\
m_{1}^{z}\left[  n\right]   &  =z_{1}\left[  n\right]  ,\\
m_{2}^{x}\left[  n\right]   &  =x_{2}\left[  n\right]  +f_{1}m_{1}^{x}\left[
n-1\right]  ,\\
m_{2}^{z}\left[  n\right]   &  =z_{2}\left[  n\right]  .
\end{align*}
Some substitutions simplify this set of recursive equations so that it becomes
the following set:%
\begin{align*}
x_{1}^{\prime}\left[  n\right]   &  =x_{1}\left[  n-1\right]  ,\\
z_{1}^{\prime}\left[  n\right]   &  =z_{1}\left[  n-1\right]  +f_{1}%
z_{2}\left[  n\right]  ,\\
x_{2}^{\prime}\left[  n\right]   &  =x_{2}\left[  n-1\right]  +f_{1}%
x_{1}\left[  n-2\right]  ,\\
z_{2}^{\prime}\left[  n\right]   &  =z_{2}\left[  n-1\right]  .
\end{align*}
We can transform this set of equations into the \textquotedblleft%
$D$-domain\textquotedblright\ with the $D$-transform \cite{book1999conv}. The
set transforms as follows:%
\begin{align*}
x_{1}^{\prime}\left(  D\right)   &  =Dx_{1}\left(  D\right)  ,\\
z_{1}^{\prime}\left(  D\right)   &  =D\left(  z_{1}\left(  D\right)
+f_{1}D^{-1}z_{2}\left(  D\right)  \right)  ,\\
x_{2}^{\prime}\left(  D\right)   &  =D\left(  x_{2}\left(  D\right)
+f_{1}Dx_{1}\left(  D\right)  \right)  ,\\
z_{2}^{\prime}\left(  D\right)   &  =Dz_{2}\left(  D\right)  .
\end{align*}
This set of transformations is linear, and we can write them as the following
matrix equation:%
\[
D\left[  \left.
\begin{array}
[c]{cc}%
1 & 0\\
f_{1}D^{-1} & 1\\
0 & 0\\
0 & 0
\end{array}
\right\vert
\begin{array}
[c]{cc}%
0 & 0\\
0 & 0\\
1 & f_{1}D\\
0 & 1
\end{array}
\right]  .
\]
The factor of $D$ accounts for the unit delay necessary to implement this
device, but it is not particularly relevant for the purposes of the
transformation (we might as well say that this quantum shift register device
implements the transformation without the factor of $D$). Postmultiplying the
vector%
\[
\left[  \left.
\begin{array}
[c]{cc}%
z_{1}\left(  D\right)  & z_{2}\left(  D\right)
\end{array}
\right\vert
\begin{array}
[c]{cc}%
x_{1}\left(  D\right)  & x_{2}\left(  D\right)
\end{array}
\right]
\]
by the above matrix gives the output vector%
\[
\left[  \left.
\begin{array}
[c]{cc}%
z_{1}^{\prime}\left(  D\right)  & z_{2}^{\prime}\left(  D\right)
\end{array}
\right\vert
\begin{array}
[c]{cc}%
x_{1}^{\prime}\left(  D\right)  & x_{2}^{\prime}\left(  D\right)
\end{array}
\right]  .
\]
The above transformation confirms our intuition concerning the propagation of
$X$ and $Z$ variables. The $D$ term on the right side of the transformation
matrix indicates that the $X$ variable of the first qubit propagates one frame
into the past with respect to itself, and the $D^{-1}$ term on the left side
of the matrix indicates that the $Z$ variable of the second qubit propagates
one frame into the future with respect to itself.%

%TCIMACRO{\FRAME{ftbpFU}{2.0358in}{1.1156in}{0pt}{\Qcb{The circuit in the above
%figure combines the circuits in Figures~\ref{fig:simple-CNOT} and
%\ref{fig:one-delay-CNOT}.}}{\Qlb{fig:one-delay-combo-CNOT}}%
%{one-delay-combo-cnot.pdf}{\special{ language "Scientific Word";
%type "GRAPHIC";  maintain-aspect-ratio TRUE;  display "USEDEF";
%valid_file "F";  width 2.0358in;  height 1.1156in;  depth 0pt;
%original-width 4.7262in;  original-height 2.5598in;  cropleft "0";
%filename 'figures/one-delay-combo-CNOT.pdf';file-properties "XNPEU";}}}%
%BeginExpansion
\begin{figure}
[ptb]
\begin{center}
\includegraphics[
natheight=2.559800in,
natwidth=4.726200in,
height=1.1156in,
width=2.0358in
]%
{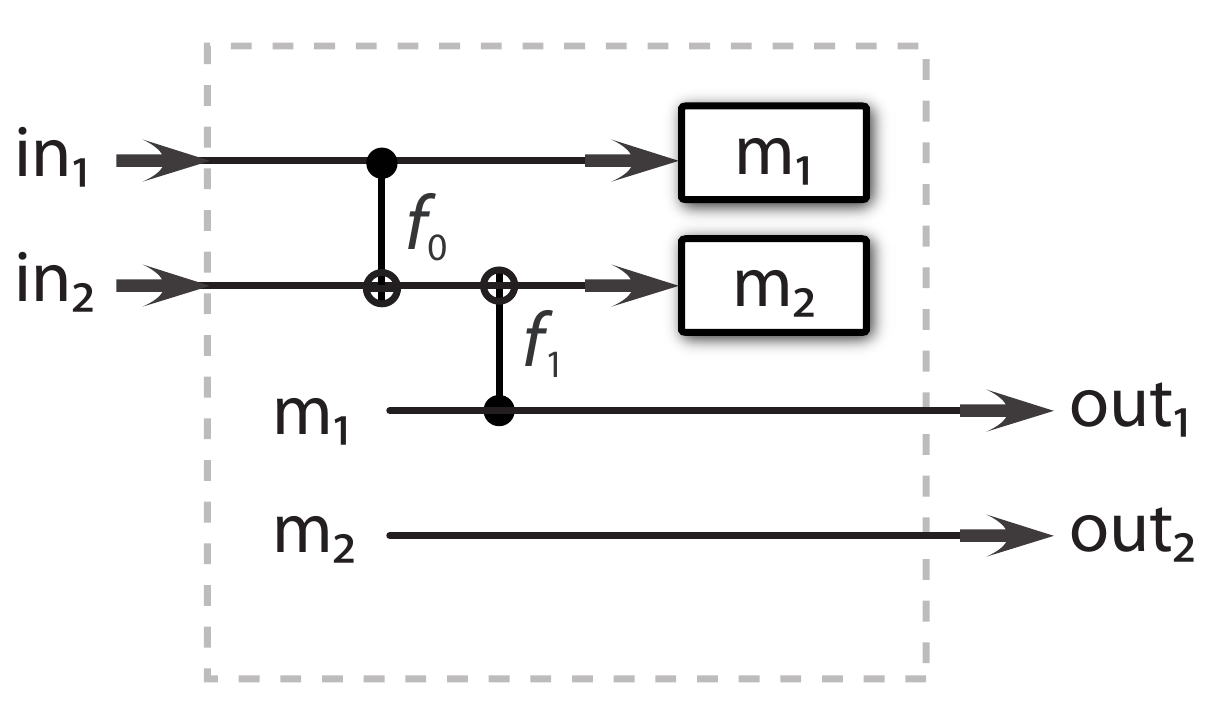}%
\caption{The circuit in the above figure combines the circuits in
Figures~\ref{fig:simple-CNOT} and \ref{fig:one-delay-CNOT}.}%
\label{fig:one-delay-combo-CNOT}%
\end{center}
\end{figure}
%EndExpansion
We now consider combining the different quantum shift register circuits
together. Suppose that we connect the outputs of the device in
Figure~\ref{fig:simple-CNOT} to the inputs of the device in
Figure~\ref{fig:one-delay-CNOT}. Figure~\ref{fig:one-delay-combo-CNOT}%
\ depicts the resulting quantum shift register circuit, and it follows that
the resulting transformation in the $D$-domain is%
\begin{equation}
D\left[  \left.
\begin{array}
[c]{cc}%
1 & 0\\
f_{0}+f_{1}D^{-1} & 1\\
0 & 0\\
0 & 0
\end{array}
\right\vert
\begin{array}
[c]{cc}%
0 & 0\\
0 & 0\\
1 & f_{0}+f_{1}D\\
0 & 1
\end{array}
\right]  . \label{eq:unit-delay-combo}%
\end{equation}

\begin{figure}
[ptb]
\begin{center}
\includegraphics[
natheight=3.246500in,
natwidth=4.726200in,
height=1.6821in,
width=2.437in
]%
{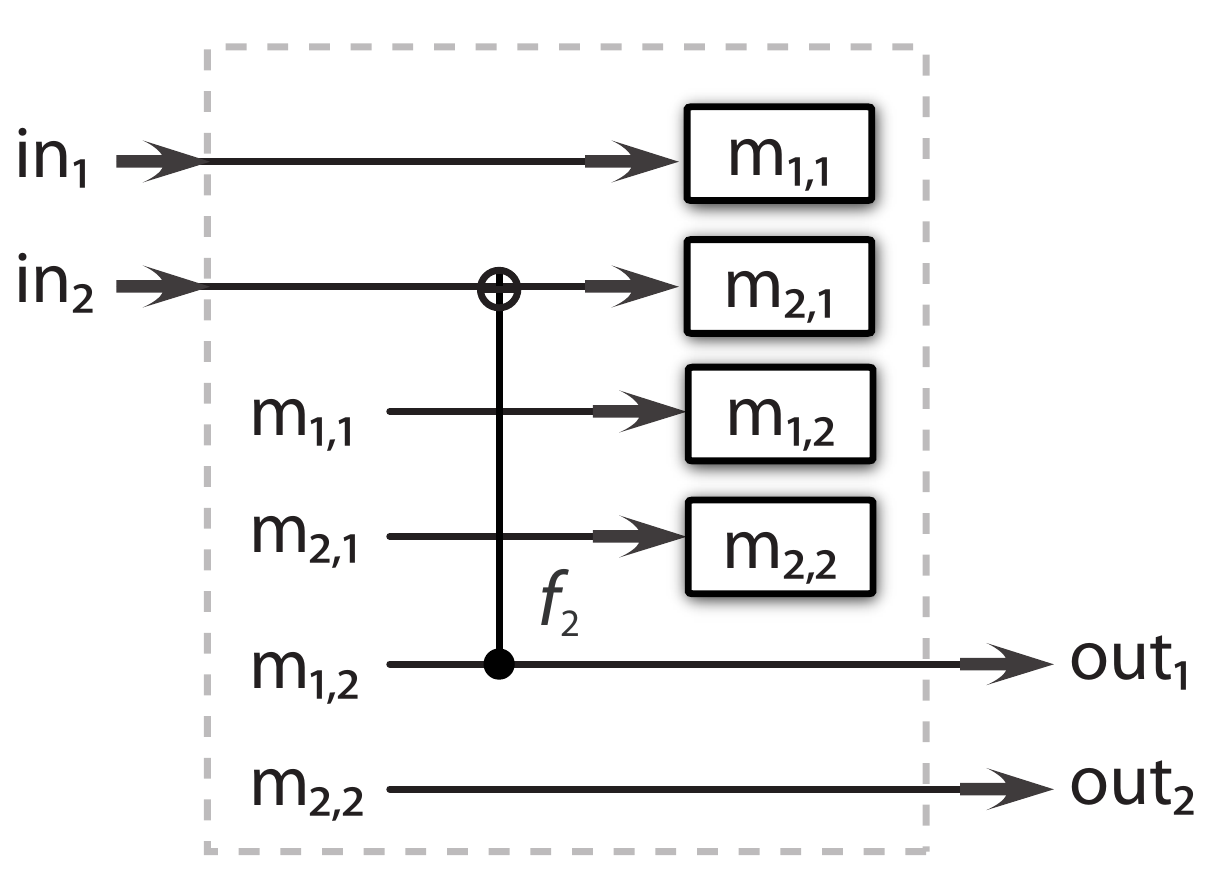}%
\caption{The circuit in the above figure implements a two-delay CNOT
transformation.}%
\label{fig:two-delay-CNOT}%
\end{center}
\end{figure}

Now consider the \textquotedblleft two-delay transformation\textquotedblright%
\ in Figure~\ref{fig:two-delay-CNOT}. The circuit is similar to the one in
Figure~\ref{fig:one-delay-CNOT}, with the exception that the first outgoing
qubit acts on the second incoming qubit and the second incoming qubit is
delayed two frames with respect to the first outgoing qubit. We now expect
that the $X$ variable propagates two frames into the past, while the $Z$
variable propagates two frames into the future. The transformation should be
as follows:%
\begin{equation}
D^{2}\left[  \left.
\begin{array}
[c]{cc}%
1 & 0\\
f_{2}D^{-1} & 1\\
0 & 0\\
0 & 0
\end{array}
\right\vert
\begin{array}
[c]{cc}%
0 & 0\\
0 & 0\\
1 & f_{2}D\\
0 & 1
\end{array}
\right]  . \label{eq:two-delay}%
\end{equation}
An analysis similar to the one for the \textquotedblleft
one-delay\textquotedblright\ CNOT transformation shows that the circuit indeed
implements the above transformation.%
%TCIMACRO{\FRAME{ftbpFU}{2.437in}{1.6821in}{0pt}{\Qcb{The circuit in the above
%figure implements a two-delay CNOT transformation.}}{\Qlb{fig:two-delay-CNOT}%
%}{two-delay-cnot.pdf}{\special{ language "Scientific Word";  type "GRAPHIC";
%width 2.437in;  height 1.6821in;  depth 0pt;  original-width 4.7262in;
%original-height 3.2465in;  cropleft "0";  croptop "1";  cropright "1";
%filename 'figures/two-delay-CNOT.pdf';file-properties "XNPEU";}}}%
%BeginExpansion
%EndExpansion

Let us connect the outputs of the device in
Figure~\ref{fig:one-delay-combo-CNOT} to the inputs of the device in
Figure~\ref{fig:two-delay-CNOT}. The resulting $D$-domain transformation
should be the multiplication of the transformation in
(\ref{eq:unit-delay-combo}) with that in (\ref{eq:two-delay}), and an analysis
with recursive equations confirms that the transformation is the following
one:%
\begin{equation}
D^{3}\left[  \left.
\begin{array}
[c]{cc}%
1 & 0\\
f_{0}+f_{1}D^{-1}+f_{2}D^{-2} & 1\\
0 & 0\\
0 & 0
\end{array}
\right\vert
\begin{array}
[c]{cc}%
0 & 0\\
0 & 0\\
1 & f_{0}+f_{1}D+f_{2}D^{2}\\
0 & 1
\end{array}
\right]  . \label{eq:two-delay-combo}%
\end{equation}

The resulting device uses three frames of memory qubits to implement the
transformation. This amount of memory seems like it may be too much,
considering that the output data only depends on the input from two frames
into the past. Is there any way to save on memory consumption?%
%TCIMACRO{\FRAME{ftbpFU}{3.3399in}{1.785in}{0pt}{\Qcb{The above circuit
%connects the outputs of the circuit in Figure~\ref{fig:one-delay-combo-CNOT}%
%\ to the inputs of the circuit in Figure~\ref{fig:two-delay-CNOT}.}%
%}{\Qlb{fig:in-between-circuit}}{in-between-combo-cnot.pdf}%
%width 3.3399in;  height 1.785in;  depth 0pt;  original-width 8.0073in;
%original-height 4.2462in;  cropleft "0";  croptop "1";  cropright "1";
%filename 'figures/in-between-combo-CNOT.pdf';file-properties "XNPEU";}}}%
%BeginExpansion
\begin{figure}
[ptb]
\begin{center}
\includegraphics[
natheight=4.246200in,
natwidth=8.007300in,
height=1.785in,
width=3.3399in
]%
{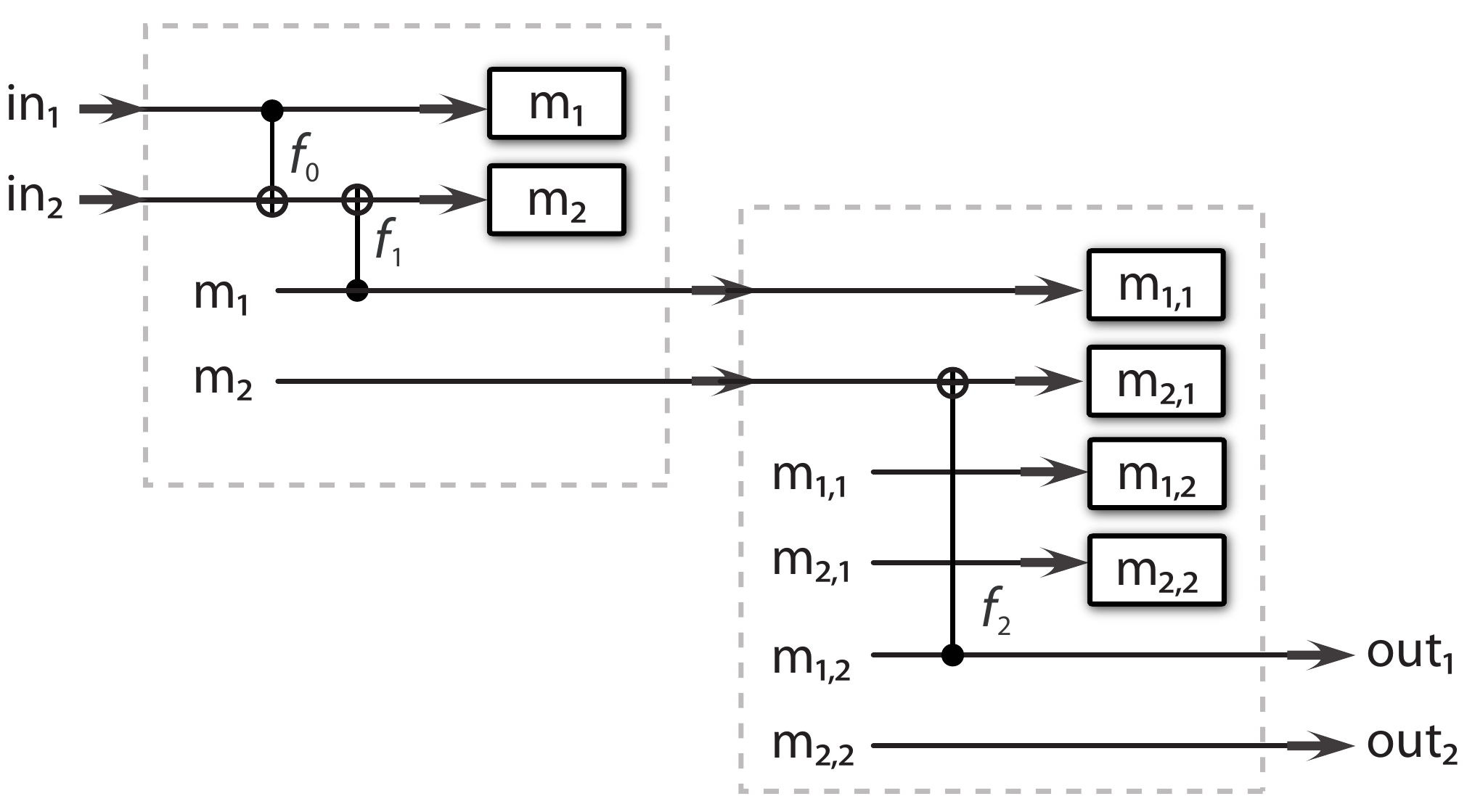}%
\caption{The above circuit connects the outputs of the circuit in
Figure~\ref{fig:one-delay-combo-CNOT}\ to the inputs of the circuit in
Figure~\ref{fig:two-delay-CNOT}.}%
\label{fig:in-between-circuit}%
\end{center}
\end{figure}
%EndExpansion

\begin{figure}
[ptb]
\begin{center}
\includegraphics[
natheight=3.246500in,
natwidth=5.193200in,
height=1.5342in,
width=2.437in
]%
{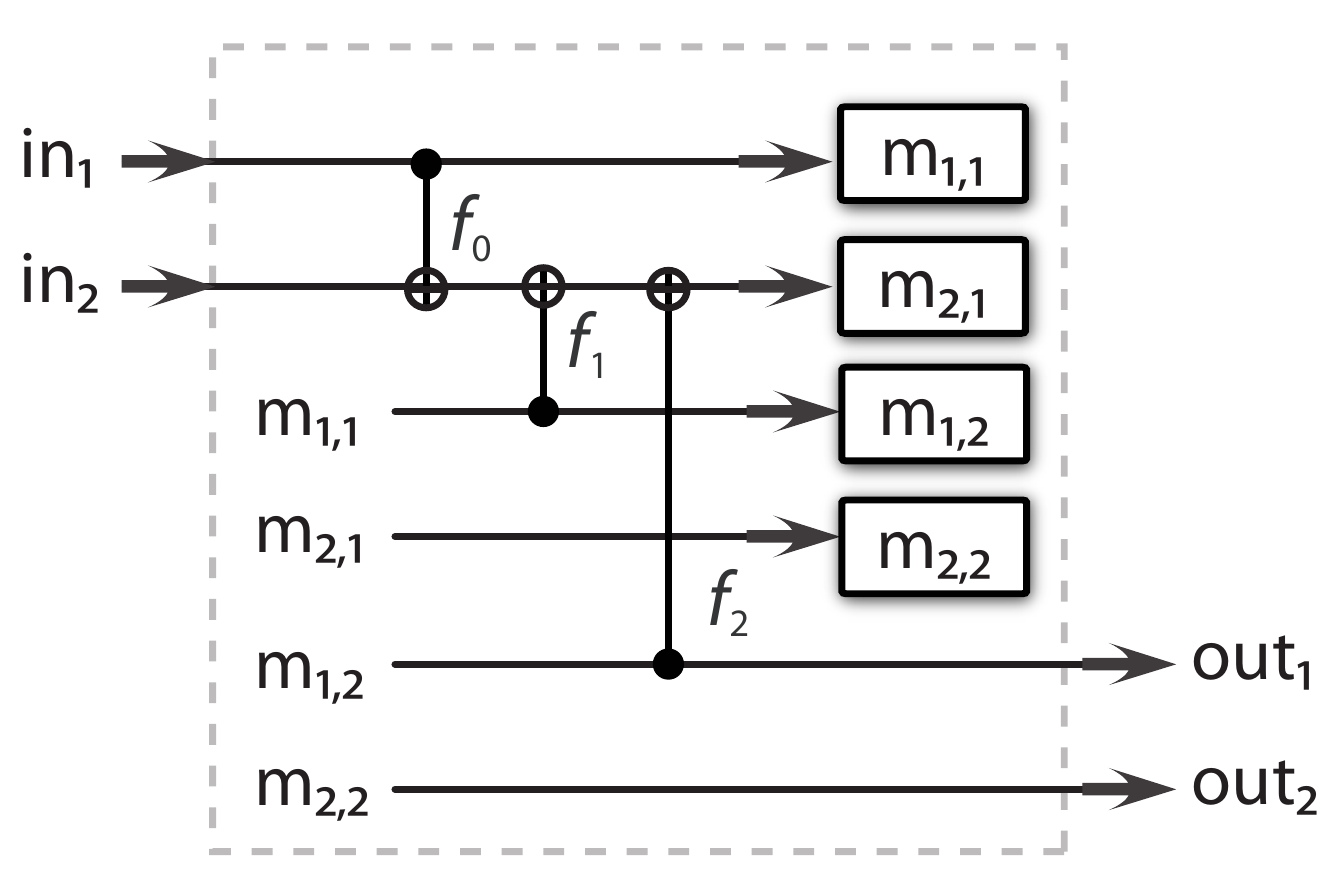}%
\caption{The above circuit reduces the amount of memory required to implement
the transformation in (\ref{eq:two-delay-combo}).}%
\label{fig:two-delay-combo-CNOT}%
\end{center}
\end{figure}

First, let us connect the outputs of the circuit in
Figure~\ref{fig:one-delay-combo-CNOT} to the inputs of the circuit in
Figure~\ref{fig:two-delay-CNOT}. Figure~\ref{fig:in-between-circuit}\ depicts
the resulting device. In this \textquotedblleft combo\textquotedblright%
\ device, the target of the CNOT\ gate conditional on $f_{2}$ does not act on
the source of the CNOT\ gate conditional on $f_{1}$. Therefore, we can commute
the \textquotedblleft$f_{2}$-gate\textquotedblright\ with the
\textquotedblleft$f_{1}$-gate.\textquotedblright\ Now, we can actually then
\textquotedblleft commute this gate through the memory\textquotedblright%
\ because it does not matter whether this CNOT\ gate acts on the qubits before
they pass through the memory or after they come out. It then follows that the
last frame of memory qubits are not necessary because there is no gate that
acts on these last qubits. Figure~\ref{fig:two-delay-combo-CNOT}\ depicts the
simplified transformation. It is also straightforward to check that the
resulting transformation is as follows:%
\begin{equation}
D^{2}\left[  \left.
\begin{array}
[c]{cc}%
1 & 0\\
f_{0}+f_{1}D^{-1}+f_{2}D^{-2} & 1\\
0 & 0\\
0 & 0
\end{array}
\right\vert
\begin{array}
[c]{cc}%
0 & 0\\
0 & 0\\
1 & f_{0}+f_{1}D+f_{2}D^{2}\\
0 & 1
\end{array}
\right]  , \label{eq:two-delay-combo-reduced}%
\end{equation}
where the premultiplying delay factor in (\ref{eq:two-delay-combo-reduced}) is
now $D^{2}$ instead of $D^{3}$ as in (\ref{eq:two-delay-combo}).%
%TCIMACRO{\FRAME{ftbpFU}{2.437in}{1.5342in}{0pt}{\Qcb{The above circuit reduces
%the amount of memory required to implement the transformation in
%(\ref{eq:two-delay-combo}).}}{\Qlb{fig:two-delay-combo-CNOT}}%
%{two-delay-combo-cnot.pdf}{\special{ language "Scientific Word";
%type "GRAPHIC";  maintain-aspect-ratio TRUE;  display "USEDEF";
%valid_file "F";  width 2.437in;  height 1.5342in;  depth 0pt;
%original-width 5.1932in;  original-height 3.2465in;  cropleft "0";
%filename 'figures/two-delay-combo-CNOT.pdf';file-properties "XNPEU";}}}%
%BeginExpansion
%EndExpansion

\section{General Encoding Algorithm}

The procedure in the previous section allows us to simplify the circuit by
eliminating the last frame of memory qubits. This procedure of determining
whether we can \textquotedblleft commute gates through the
memory\textquotedblright\ is a general one that we can employ for reducing
memory in quantum shift register circuits. In the above example, we can
determine the number of frames of memory that are necessary by considering the
absolute degree of the polynomial transformation in (\ref{eq:two-delay-combo})
(without including the $D^{3}$ prefactor). The \textit{absolute degree}
$\left\vert \deg\right\vert \left(  B\left(  D\right)  \right)  $ of a
polynomial matrix $B\left(  D\right)  $ is%
\[
\left\vert \deg\right\vert \left(  B\left(  D\right)  \right)  \equiv
\max\left\{  d_{1},d_{2}\right\}  ,
\]
where%
\begin{align*}
d_{1} &  \equiv\max_{i,j}\left\{  \deg\left(  \left[  B\left(  D\right)
\right]  _{ij}\right)  \right\}  ,\\
d_{2} &  \equiv\max_{i,j}\left\{  \left\vert \text{del}\left(  \left[
B\left(  D\right)  \right]  _{ij}\right)  \right\vert \right\}  ,
\end{align*}
del$\left(  b\left(  D\right)  \right)  $ is the lowest power in the
polynomial $b\left(  D\right)  $, and the absolute degree is modulo any
prefactor terms such as the $D^{3}$ in (\ref{eq:two-delay-combo}). In the case
of the transformation in (\ref{eq:two-delay-combo}), the absolute degree is
equal to two, so we should expect to have two frames of memory qubits.
Theorem~\ref{thm:memory-CSS}\ generalizes this idea by showing that the
absolute degree of an encoding matrix corresponds to the amount of memory that
a CSS\ quantum convolutional code requires.

The procedure in the previous section demonstrates a general procedure for
constructing quantum shift register circuits for quantum convolutional codes.
We can break the encoding operation into elementary operations as the
Grassl-R\"{o}tteler encoding algorithm
does~\cite{ieee2006grassl,isit2006grassl,ieee2007grassl,arx2007wildeEAQCC}.
The general procedure implements each elementary operation with a quantum
shift register circuit, connects the outputs of one quantum shift register
circuit to the inputs of the next, and determines if it is possible to
\textquotedblleft commute gates through memory\textquotedblright\ as shown in
the above example. This procedure produces a quantum shift register encoding
circuit that uses the minimal amount of memory.

\section{Example of a Quantum Shift Register Encoding Circuit for a
CSS\ Quantum Convolutional Code}

\begin{figure}
[ptb]
\begin{center}
\includegraphics[
natheight=10.400200in,
natwidth=12.174000in,
height=2.1326in,
width=3.4402in
]%
{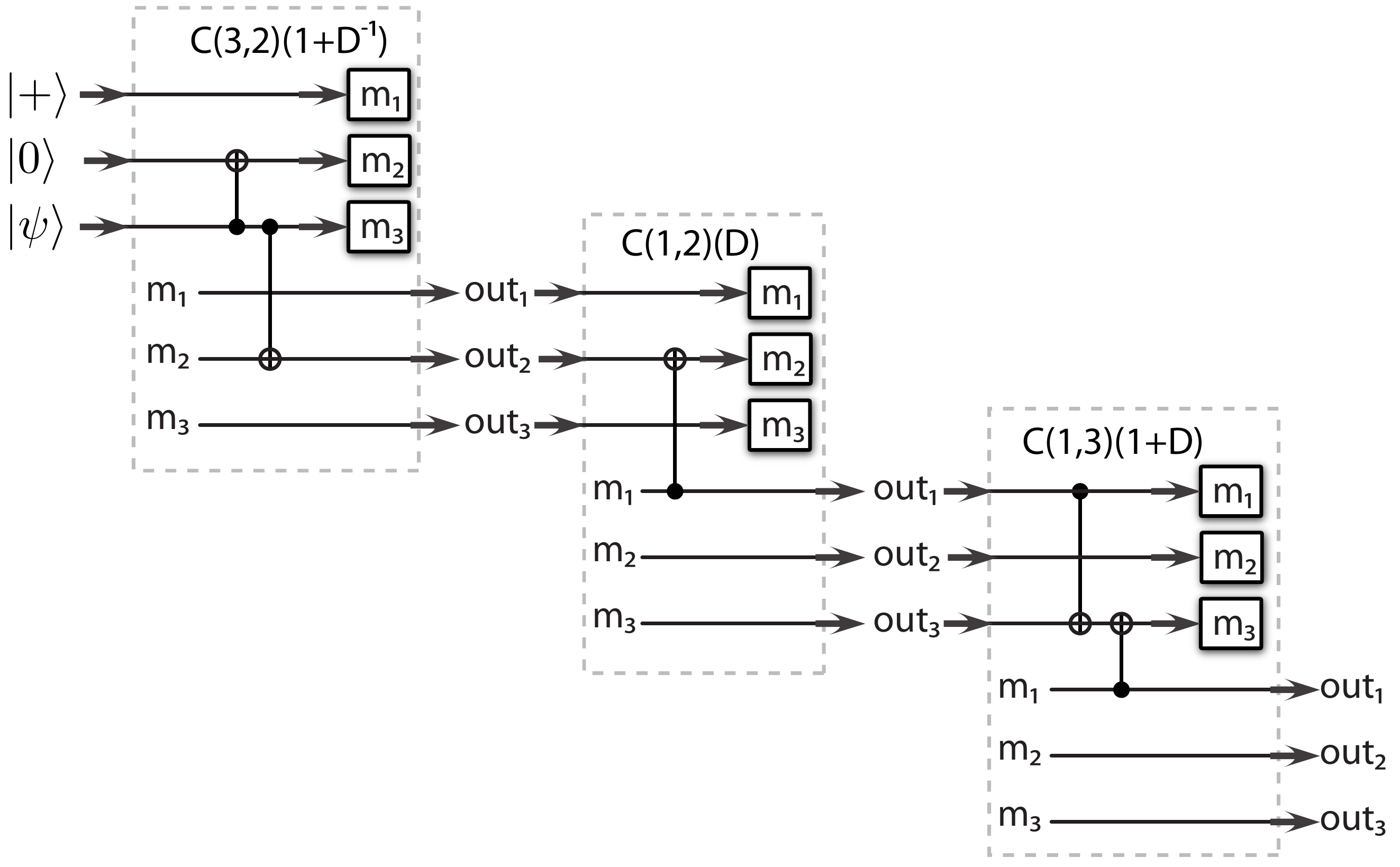}%
\caption{The above circuit implements the set of transformations outlined in
(\ref{eq:first-CNOTs}-\ref{eq:second-CNOTs}).}%
\label{fig:CSS-full-circuit}%
\end{center}
\end{figure}

Let us consider a simple example of a CSS\ quantum convolutional code
\cite{PhysRevA.54.1098,PhysRevLett.77.793}. Its stabilizer matrix
\cite{arxiv2004olliv,isit2006grassl} is as follows:%
\begin{equation}
\left[  \left.
\begin{array}
[c]{ccc}%
0 & 0 & 0\\
D & 1 & 1+D
\end{array}
\right\vert
\begin{array}
[c]{ccc}%
1 & D & 1+D\\
0 & 0 & 0
\end{array}
\right]  . \label{eq:simple-example-CSS-code}%
\end{equation}
I now show how to encode the above quantum convolutional code using a slight
modification of the Grassl-R\"{o}tteler encoding algorithm for CSS\ codes
\cite{ieee2007grassl}. One begins with the stabilizer matrix for two ancilla
qubits per frame:%
\[
\left[  \left.
\begin{array}
[c]{ccc}%
0 & 0 & 0\\
0 & 1 & 0
\end{array}
\right\vert
\begin{array}
[c]{ccc}%
1 & 0 & 0\\
0 & 0 & 0
\end{array}
\right]  .
\]
The first ancilla qubit of every frame is in the state $\left\vert
+\right\rangle $ and the second ancilla qubit of every frame is in the state
$\left\vert 0\right\rangle $. First send the three qubits through a quantum
shift register device that implements a\ CNOT$\left(  3,2\right)  \left(
1+D^{-1}\right)  $. This notation indicates that there is a CNOT\ gate from
the third qubit to the second in the same frame and to the second in a future
frame. The stabilizer becomes%
\begin{equation}
\left[  \left.
\begin{array}
[c]{ccc}%
0 & 0 & 0\\
0 & 1 & 1+D
\end{array}
\right\vert
\begin{array}
[c]{ccc}%
1 & 0 & 0\\
0 & 0 & 0
\end{array}
\right]  . \label{eq:first-CNOTs}%
\end{equation}
Then send the three qubits through a quantum shift register device that
performs a CNOT$\left(  1,2\right)  \left(  D\right)  $ (indicating a
CNOT\ from the first qubit in one frame to the second in a delayed frame) and
another quantum shift register device that performs a CNOT$\left(  1,3\right)
\left(  1+D\right)  $. The stabilizer becomes%
\begin{equation}
\left[  \left.
\begin{array}
[c]{ccc}%
0 & 0 & 0\\
D & 1 & 1+D
\end{array}
\right\vert
\begin{array}
[c]{ccc}%
1 & D & 1+D\\
0 & 0 & 0
\end{array}
\right]  , \label{eq:second-CNOTs}%
\end{equation}
and is now encoded. Note that the above circuit is a \textquotedblleft
classical\textquotedblright\ circuit in the sense that it uses only CNOT gates
in its implementation. Figure~\ref{fig:CSS-full-circuit}\ depicts the quantum
shift register circuit corresponding to the above operations.%
%TCIMACRO{\FRAME{ftbpFU}{3.4402in}{2.1326in}{0pt}{\Qcb{The above circuit
%implements the set of transformations outlined in (\ref{eq:first-CNOTs}%
%-\ref{eq:second-CNOTs}).}}{\Qlb{fig:CSS-full-circuit}}{css-full-circuit.pdf}%
%width 3.4402in;  height 2.1326in;  depth 0pt;  original-width 12.174in;
%original-height 10.4002in;  cropleft "0";  croptop "1";  cropright "1";
%filename 'figures/CSS-full-circuit.pdf';file-properties "XNPEU";}}}%
%BeginExpansion
%EndExpansion

\begin{figure}
[ptb]
\begin{center}
\includegraphics[
natheight=6.447200in,
natwidth=12.333100in,
height=1.9873in,
width=3.4402in
]%
{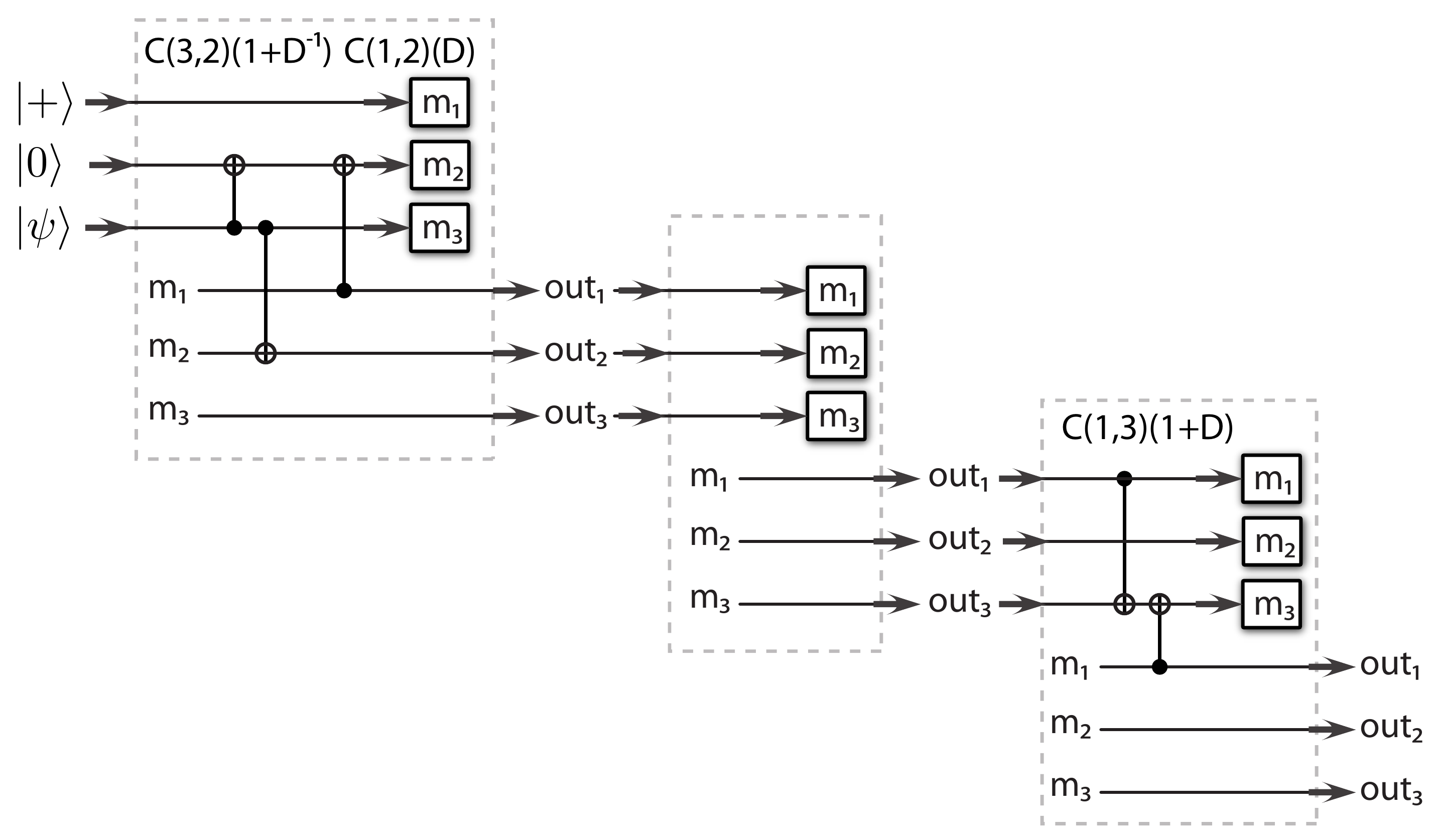}%
\caption{The above figure depicts a simplification of the circuit in
Figure~\ref{fig:CSS-full-circuit}.}%
\label{fig:CSS-mod-circuit}%
\end{center}
\end{figure}

\begin{figure}
[ptb]
\begin{center}
\includegraphics[
natheight=4.946700in,
natwidth=9.020000in,
height=1.9121in,
width=3.3399in
]%
{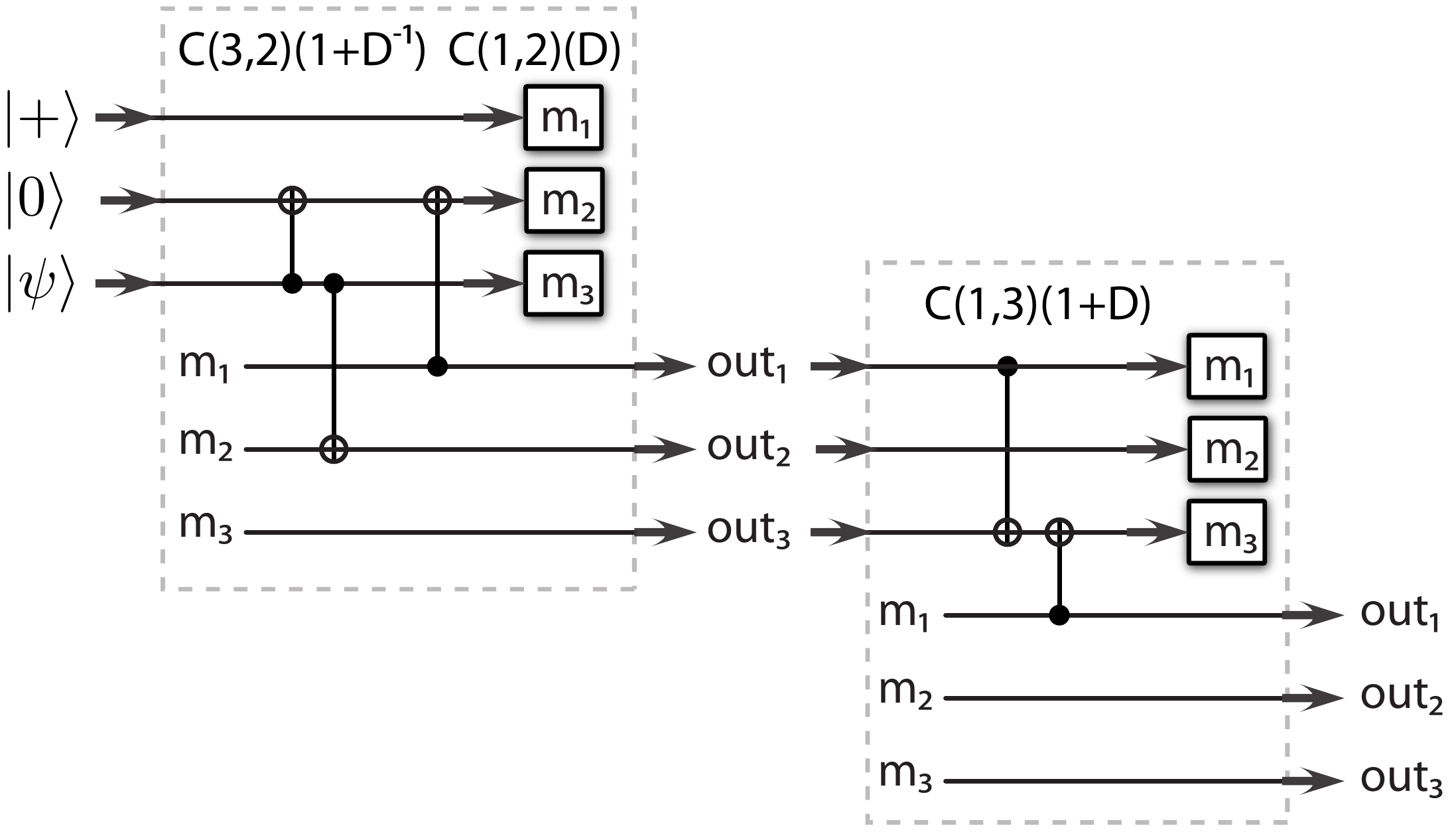}%
\caption{The above figure depicts a simplified version of the circuit in
Figure~\ref{fig:CSS-mod-circuit} where we have removed the second unnecessary
quantum shift register circuit. The above circuit uses less memory than the
one in Figure~\ref{fig:CSS-mod-circuit}, while still effecting the same
transformation.}%
\label{fig:CSS-final-mod-circuit}%
\end{center}
\end{figure}

\begin{figure}
[ptb]
\begin{center}
\includegraphics[
natheight=3.586400in,
natwidth=6.634000in,
height=1.7755in,
width=3.3399in
]%
{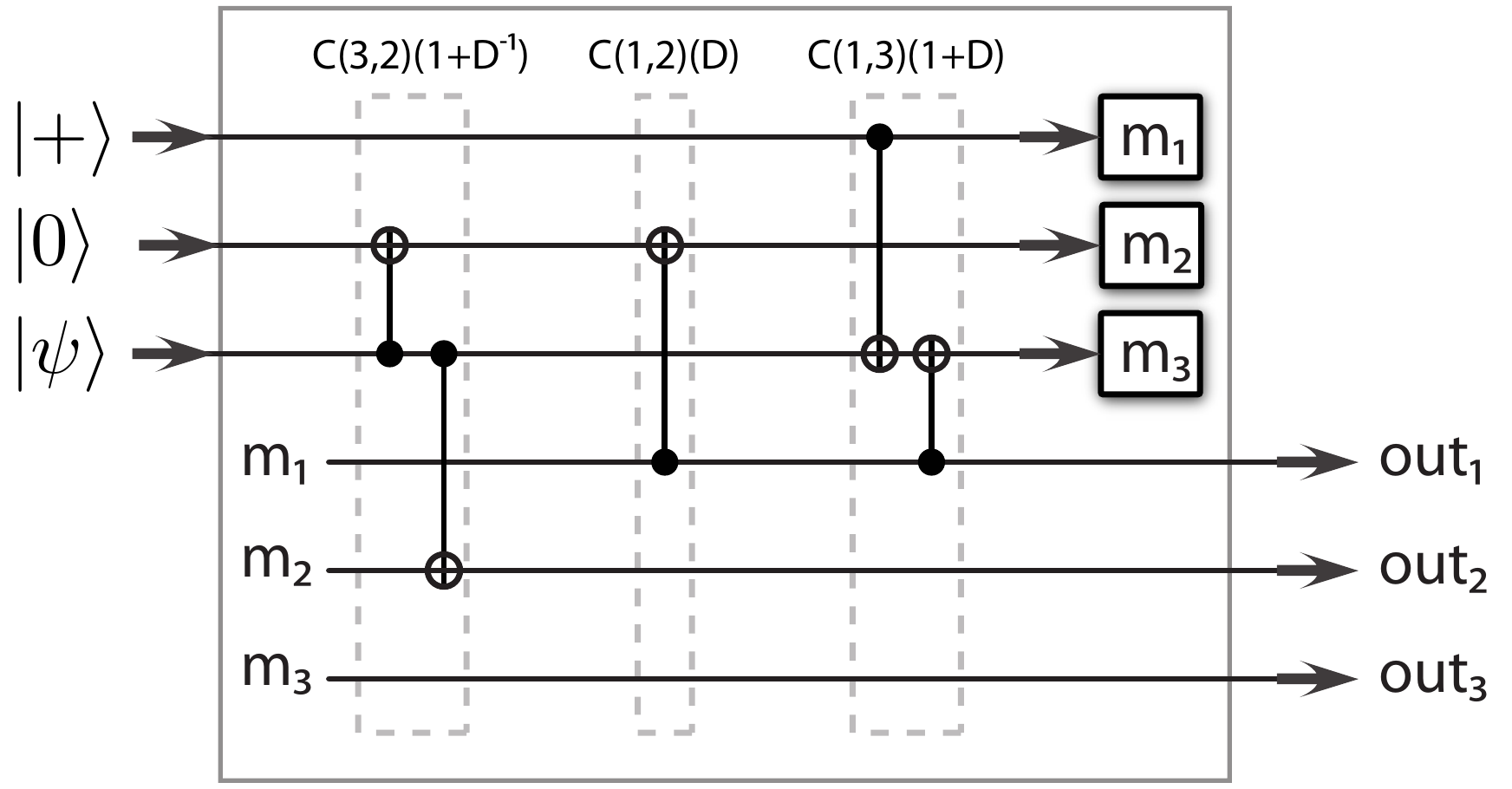}%
\caption{The circuit in the above figure is a quantum shift register encoding
circuit for the CSS quantum convolutional code in
(\ref{eq:simple-example-CSS-code}). }%
\label{fig:CSS-circuit}%
\end{center}
\end{figure}

It again seems that the circuit in Figure~\ref{fig:CSS-full-circuit} is
wasteful in memory consumption. Is there anything we can do to simplify this
circuit? First notice that the target qubit of the CNOT\ gate in the second
quantum shift register is the same as the target qubit of the second
CNOT\ gate in the first quantum shift register. It follows that these two
gates commute so that we can act with the CNOT\ gate in the second quantum
shift register before acting with the second CNOT\ gate of the first quantum
shift register. But we can do even better. Acting first with the CNOT\ gate in
the second quantum shift register is equivalent to having it act before the
first frame of memory qubits gets delayed. Figure~\ref{fig:CSS-mod-circuit}%
\ depicts this simplification.
%TCIMACRO{\FRAME{ftbpFU}{3.4402in}{1.9873in}{0pt}{\Qcb{The above figure depicts
%a simplification of the circuit in Figure~\ref{fig:CSS-full-circuit}.}%
%}{\Qlb{fig:CSS-mod-circuit}}{css-mod-circuit.pdf}%
%width 3.4402in;  height 1.9873in;  depth 0pt;  original-width 12.3331in;
%original-height 6.4472in;  cropleft "0";  croptop "1";  cropright "1";
%filename 'figures/CSS-mod-circuit.pdf';file-properties "XNPEU";}}}%
%BeginExpansion
%EndExpansion
But glancing at Figure~\ref{fig:CSS-mod-circuit}, it is now clear that the
second quantum shift register circuit no longer serves any purpose. We may
remove it from the circuit. Figure~\ref{fig:CSS-final-mod-circuit}\ displays
the resulting simplified circuit.%
%TCIMACRO{\FRAME{ftbpFU}{3.3399in}{1.9121in}{0pt}{\Qcb{The above figure depicts
%a simplified version of the circuit in Figure~\ref{fig:CSS-mod-circuit} where
%we have removed the second unnecessary quantum shift register circuit. The
%above circuit uses less memory than the one in
%Figure~\ref{fig:CSS-mod-circuit}, while still effecting the same
%transformation.}}{\Qlb{fig:CSS-final-mod-circuit}}{css-final-mod-circuit.pdf}%
%width 3.3399in;  height 1.9121in;  depth 0pt;  original-width 9.02in;
%original-height 4.9467in;  cropleft "0";  croptop "1";  cropright "1";
%filename 'figures/CSS-final-mod-circuit.pdf';file-properties "XNPEU";}}}%
%BeginExpansion
%EndExpansion
We can apply a similar logic to the two gates in the second quantum shift
register of Figure~\ref{fig:CSS-final-mod-circuit} because the two gates there
commute with the preceding gates. Performing a similar simplification and
elimination of the last frame of memory qubits leads to the final circuit.
Figure~\ref{fig:CSS-circuit}\ depicts the quantum shift register circuit that
encodes this quantum convolutional code with one frame of memory qubits.%
%TCIMACRO{\FRAME{ftbpFU}{3.3399in}{1.7755in}{0pt}{\Qcb{The circuit in the above
%figure is a quantum shift register encoding circuit for the CSS quantum
%convolutional code in (\ref{eq:simple-example-CSS-code}). }}%
%{\Qlb{fig:CSS-circuit}}{css-circuit.pdf}%
%width 3.3399in;  height 1.7755in;  depth 0pt;  original-width 6.634in;
%original-height 3.5864in;  cropleft "0";  croptop "1";  cropright "1";
%cropbottom "0";  filename 'figures/CSS-circuit.pdf';file-properties "XNPEU";}%
%}}%
%BeginExpansion
%EndExpansion

The overall encoding matrix for this code is%
\begin{align*}
&  \text{CNOT}\left(  3,2\right)  \left(  1+D^{-1}\right)  \text{CNOT}\left(
1,2\right)  \left(  D\right)  \text{CNOT}\left(  1,3\right)  \left(
1+D\right) \\
&  =\left[  \left.
\begin{array}
[c]{ccc}%
1 & 0 & 0\\
D & 1 & D+1\\
D^{-1}+1 & 0 & 1\\
0 & 0 & 0\\
0 & 0 & 0\\
0 & 0 & 0
\end{array}
\right\vert
\begin{array}
[c]{ccc}%
0 & 0 & 0\\
0 & 0 & 0\\
0 & 0 & 0\\
1 & D & D+1\\
0 & 1 & 0\\
0 & D^{-1}+1 & 1
\end{array}
\right]  .
\end{align*}
The absolute degree of the encoding matrix is one, and thus, this CSS\ code
requires one frame of memory qubits. Theorem~\ref{thm:memory-CSS} generalizes
this result to show that the memory of the encoding circuit for any
CSS\ quantum convolutional code is given by the absolute degree of the
encoding matrix for the circuit.

\section{Primitive Quantum Shift Register Circuits for CSS\ Quantum
Convolutional Codes}

\label{sec:finite-depth-CNOTs}In this section, I outline some basic primitive
operations that are useful building blocks for the quantum shift register
circuits of CSS\ (and non-CSS) quantum convolutional codes. I illustrate delay
elements and finite-depth CNOT\ operations.

\subsection{Delay Operations}

\begin{figure}
[ptb]
\begin{center}
\includegraphics[
natheight=0.920200in,
natwidth=3.166900in,
height=0.4869in,
width=1.612in
]%
{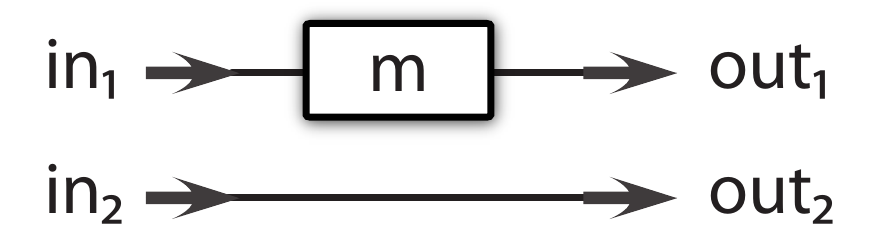}%
\caption{A circuit that implements a simple delay operation on the first qubit
in each frame.}%
\label{fig:delay}%
\end{center}
\end{figure}

The simplest operation that we can perform with a quantum shift register
circuit is to delay one qubit with respect to the others in a given frame. The
way to implement this operation is simply to insert a memory element on the
qubit that we wish to delay. Figure~\ref{fig:delay}\ depicts this delay
operation. Suppose that the Pauli operators for the two qubits in the example
are as follows (with the convention that the \textquotedblleft
Z\textquotedblright\ operators are on the left and the \textquotedblleft
X\textquotedblright\ operators are on the right):%
%TCIMACRO{\FRAME{ftbpFU}{1.612in}{0.4869in}{0pt}{\Qcb{A circuit that implements
%a simple delay operation on the first qubit in each frame.}}{\Qlb{fig:delay}%
%}{delay.pdf}{\special{ language "Scientific Word";  type "GRAPHIC";
%width 1.612in;  height 0.4869in;  depth 0pt;  original-width 3.1669in;
%original-height 0.9202in;  cropleft "0";  croptop "1";  cropright "1";
%cropbottom "0";  filename 'figures/delay.pdf';file-properties "XNPEU";}}}%
%BeginExpansion
%EndExpansion%
\[
\left[  \left.
\begin{array}
[c]{cc}%
1 & 0\\
0 & 1\\
0 & 0\\
0 & 0
\end{array}
\right\vert
\begin{array}
[c]{cc}%
0 & 0\\
0 & 0\\
1 & 0\\
0 & 1
\end{array}
\right]  .
\]
The circuit in Figure~\ref{fig:delay}\ transforms the operators as follows:%
\[
\left[  \left.
\begin{array}
[c]{cc}%
D & 0\\
0 & 1\\
0 & 0\\
0 & 0
\end{array}
\right\vert
\begin{array}
[c]{cc}%
0 & 0\\
0 & 0\\
D & 0\\
0 & 1
\end{array}
\right]  .
\]

\subsection{Building Finite-Depth CNOT\ Operations}

I now show how to generalize the above examples to implement a general
CNOT\ finite-depth operation. Suppose that we have two qubits on which we
would like to perform a finite-depth operation \footnote{A finite-depth
operation is one that takes any finite weight Pauli operator to another
finite-weight Pauli operator.}. The Pauli operators for these qubits are as
follows:%
\[
\left[  \left.
\begin{array}
[c]{cc}%
1 & 0\\
0 & 1\\
0 & 0\\
0 & 0
\end{array}
\right\vert
\begin{array}
[c]{cc}%
0 & 0\\
0 & 0\\
1 & 0\\
0 & 1
\end{array}
\right]  .
\]
A general shift-invariant finite-depth CNOT\ operation translates the above
set of operators to the following set:%
\begin{equation}
\left[  \left.
\begin{array}
[c]{cc}%
1 & 0\\
f\left(  D^{-1}\right)  & 1\\
0 & 0\\
0 & 0
\end{array}
\right\vert
\begin{array}
[c]{cc}%
0 & 0\\
0 & 0\\
1 & f\left(  D\right) \\
0 & 1
\end{array}
\right]  , \label{eq:finite-depth-poly-transform}%
\end{equation}
where $f\left(  D\right)  $ is some arbitrary binary polynomial:%
\[
f\left(  D\right)  =\sum_{i=0}^{M}f_{i}D^{i}.
\]

\begin{theorem}
\label{thm:finite-depth-x}The circuit in Figure~\ref{fig:CNOT-finite-depth}%
\ implements the transformation in (\ref{eq:finite-depth-poly-transform}) and
it requires $M$ frames of memory qubits.
\end{theorem}

\begin{proof}
The proof of this theorem uses linear system theoretic techniques by
considering symplectic binary vectors that correspond to the Pauli operators
for the incoming qubits, the outgoing ones, and the memory qubits. We can
formulate a system of recursive equations involving these binary variables
similar to how we did for the previous examples. Let us label the bit
representations of the $X$ Pauli operators for all the qubits as follows:%
\[
x_{1}^{\prime},x_{2}^{\prime},m_{1,1}^{x},m_{2,1}^{x},m_{1,2}^{x},m_{2,2}%
^{x},\ldots,m_{1,M}^{x},m_{2,M}^{x},x_{1},x_{2},
\]
where the primed variables are the outputs and the unprimed are the inputs.
Let us label the bit representations of the $Z$ Pauli operators similarly:%
\[
z_{1}^{\prime},z_{2}^{\prime},m_{1,1}^{z},m_{2,1}^{z},m_{1,2}^{z},m_{2,2}%
^{z},\ldots,m_{1,M}^{z},m_{2,M}^{z},z_{1},z_{2}.
\]%
%TCIMACRO{\FRAME{ftbpFU}{3.039in}{2.284in}{0pt}{\Qcb{The circuit in the above
%figure implements the transformation in (\ref{eq:finite-depth-poly-transform}%
%). }}{\Qlb{fig:CNOT-finite-depth}}{cnot-finite-depth.pdf}%
%width 3.039in;  height 2.284in;  depth 0pt;  original-width 6.5397in;
%original-height 4.9in;  cropleft "0";  croptop "1";  cropright "1";
%filename 'figures/CNOT-finite-depth.pdf';file-properties "XNPEU";}}}%
%BeginExpansion
\begin{figure}
[ptb]
\begin{center}
\includegraphics[
natheight=4.900000in,
natwidth=6.539700in,
height=2.284in,
width=3.039in
]%
{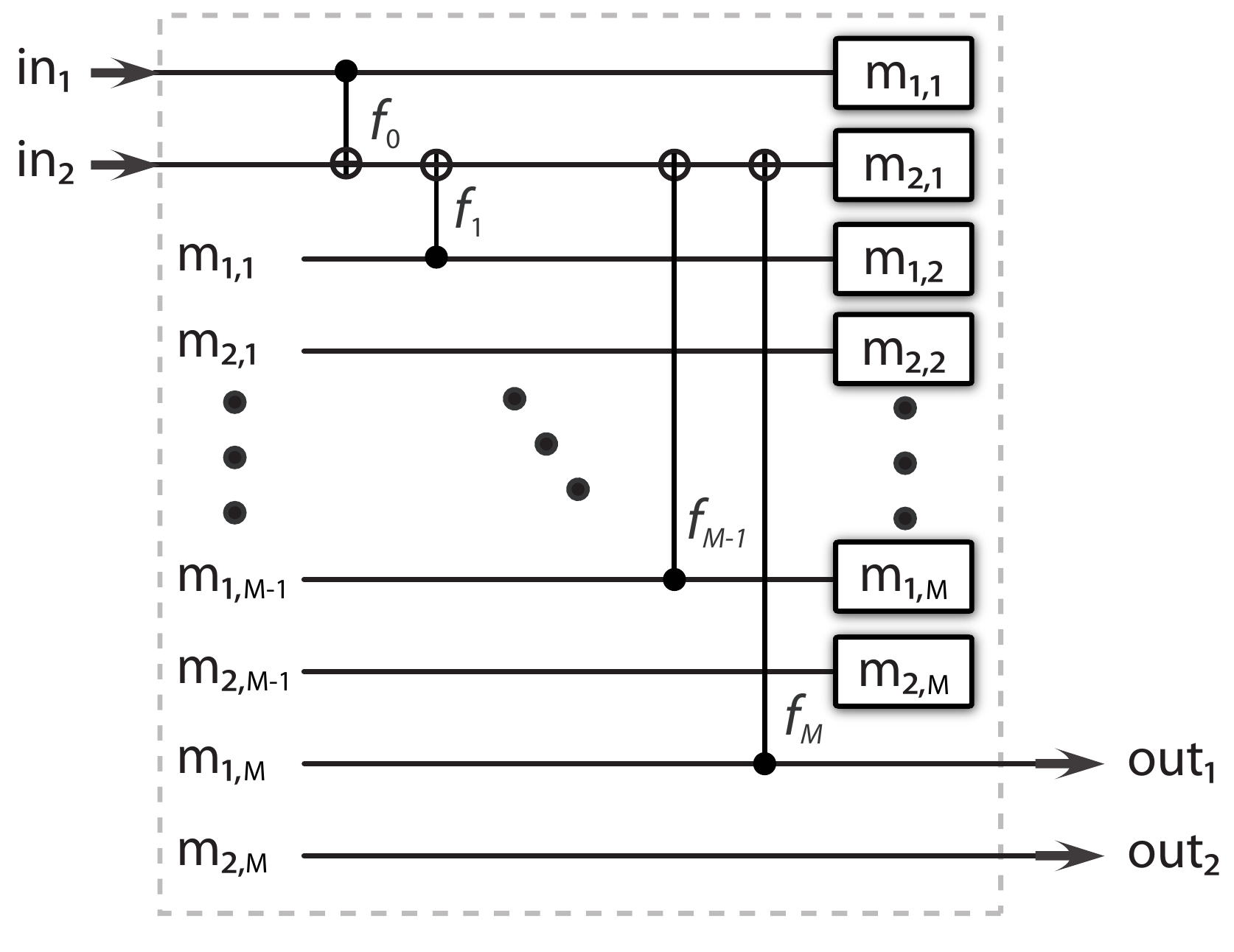}%
\caption{The circuit in the above figure implements the transformation in
(\ref{eq:finite-depth-poly-transform}). }%
\label{fig:CNOT-finite-depth}%
\end{center}
\end{figure}
%EndExpansion
The circuit in Figure~\ref{fig:CNOT-finite-depth} implements the following set
of recursive \textquotedblleft X\textquotedblright\ equations:%
\begin{align*}
x_{1}^{\prime}\left[  n\right]   &  =m_{1,M}^{x}\left[  n-1\right]  ,\\
x_{2}^{\prime}\left[  n\right]   &  =m_{2,M}^{x}\left[  n-1\right]  ,\\
m_{1,1}^{x}\left[  n\right]   &  =x_{1}\left[  n\right]  ,\\
m_{2,1}^{x}\left[  n\right]   &  =x_{2}\left[  n\right]  +f_{0}x_{1}\left[
n\right]  +\sum_{i=1}^{M}f_{i}m_{1,i}^{x}\left[  n-1\right]  ,
\end{align*}
and $\forall i=2\ldots M$,%
\begin{align*}
m_{1,i}^{x}\left[  n\right]   &  =m_{1,i-1}^{x}\left[  n-1\right]  ,\\
m_{2,i}^{x}\left[  n\right]   &  =m_{2,i-1}^{x}\left[  n-1\right]  .
\end{align*}
The set of \textquotedblleft Z\textquotedblright\ recursive equations is as
follows:%
\begin{align*}
z_{1}^{\prime}\left[  n\right]   &  =m_{1,M}^{z}\left[  n-1\right]
+f_{M}\ z_{2}\left[  n\right]  ,\\
z_{2}^{\prime}\left[  n\right]   &  =m_{2,M}^{z}\left[  n-1\right]  ,\\
m_{1,1}^{z}\left[  n\right]   &  =z_{1}\left[  n\right]  +f_{0}\ z_{2}\left[
n\right]  ,\\
m_{2,1}^{z}\left[  n\right]   &  =z_{2}\left[  n\right]  ,
\end{align*}
and $\forall i=2,\ldots,M$,%
\begin{align*}
m_{1,i}^{z}\left[  n\right]   &  =m_{1,i-1}^{z}\left[  n-1\right]
+f_{i-1}\ z_{2}\left[  n\right]  ,\\
m_{2,i}^{z}\left[  n\right]   &  =m_{2,i-1}^{z}\left[  n-1\right]  .
\end{align*}
Simplifying\ the \textquotedblleft X\textquotedblright\ equations gives the
following two equations:%
\begin{align*}
x_{1}^{\prime}\left[  n\right]   &  =x_{1}\left[  n-M\right]  ,\\
x_{2}^{\prime}\left[  n\right]   &  =x_{2}\left[  n-M\right]  +\sum_{i=0}%
^{M}f_{i}x_{1}\left[  n-M-i\right]  .
\end{align*}
Simplifying\ the \textquotedblleft Z\textquotedblright\ equations gives the
following two equations:%
\begin{align*}
z_{1}^{\prime}\left[  n\right]   &  =z_{1}\left[  n-M\right]  +\sum_{i=0}%
^{M}f_{i}z_{2}\left[  n-M+i\right]  ,\\
z_{2}^{\prime}\left[  n\right]   &  =z_{2}\left[  n-M\right]  .
\end{align*}
Applying the $D$-transform to the above gives the following set of equations:%
\begin{align*}
x_{1}^{\prime}\left(  D\right)   &  =D^{M}x_{1}\left(  D\right)  ,\\
x_{2}^{\prime}\left(  D\right)   &  =D^{M}\left(  x_{2}\left(  D\right)
+\sum_{i=0}^{M}f_{i}D^{i}x_{1}\left(  D\right)  \right) \\
&  =D^{M}\left(  x_{2}\left(  D\right)  +f\left(  D\right)  x_{1}\left(
D\right)  \right)  ,\\
z_{1}^{\prime}\left(  D\right)   &  =D^{M}\left(  z_{1}\left(  D\right)
+\sum_{i=0}^{M}f_{i}D^{-i}z_{2}\left(  D\right)  \right) \\
&  =D^{M}\left(  z_{1}\left(  D\right)  +f\left(  D^{-1}\right)  z_{2}\left(
D\right)  \right)  ,\\
z_{2}^{\prime}\left(  D\right)   &  =D^{M}z_{2}\left(  D\right)  .
\end{align*}
Rewriting the above set of equations as a matrix transformation reveals that
it is equivalent to the transformation in
(\ref{eq:finite-depth-poly-transform}) (modulo the factor $D^{M}$):%
\[
\left[  \left.
\begin{array}
[c]{cc}%
1 & 0\\
f\left(  D^{-1}\right)  & 1\\
0 & 0\\
0 & 0
\end{array}
\right\vert
\begin{array}
[c]{cc}%
0 & 0\\
0 & 0\\
1 & f\left(  D\right) \\
0 & 1
\end{array}
\right]  D^{M}.
\]
Postmultiplying the following vector by the above transformation%
\[
\left[  \left.
\begin{array}
[c]{cc}%
z_{1}\left(  D\right)  & z_{2}\left(  D\right)
\end{array}
\right\vert
\begin{array}
[c]{cc}%
x_{1}\left(  D\right)  & x_{2}\left(  D\right)
\end{array}
\right]  ,
\]
gives the following output vector:%
\[
\left[  \left.
\begin{array}
[c]{cc}%
z_{1}^{\prime}\left(  D\right)  & z_{2}^{\prime}\left(  D\right)
\end{array}
\right\vert
\begin{array}
[c]{cc}%
x_{1}^{\prime}\left(  D\right)  & x_{2}^{\prime}\left(  D\right)
\end{array}
\right]  .
\]
The circuit in Figure~\ref{fig:CNOT-finite-depth} uses $M$ frames of memory
qubits ($2M$ actual memory qubits).
\end{proof}

Suppose now that we reverse the direction of the CNOT\ gates in
Figure~\ref{fig:CNOT-finite-depth}. The result is to perform a shift-invariant
finite-depth CNOT\ operation \textquotedblleft conjugate\textquotedblright\ to
that in (\ref{eq:finite-depth-poly-transform}):%
\begin{equation}
\left[  \left.
\begin{array}
[c]{cc}%
1 & f\left(  D\right) \\
0 & 1\\
0 & 0\\
0 & 0
\end{array}
\right\vert
\begin{array}
[c]{cc}%
0 & 0\\
0 & 0\\
1 & 0\\
f\left(  D^{-1}\right)  & 1
\end{array}
\right]  . \label{eq:finite-depth-poly-transform-conj}%
\end{equation}
It merely switches the roles of the $X$ and $Z$ variables.%
%TCIMACRO{\FRAME{ftbpFU}{3.2993in}{2.4794in}{0pt}{\Qcb{The circuit in the above
%figure implements the transformation in
%(\ref{eq:finite-depth-poly-transform-conj}). Comparing the circuit in the
%above figure to the one in Figure~\ref{fig:CNOT-finite-depth} reveals that it
%merely flips the direction of the CNOT gates.}}%
%{\Qlb{fig:CNOT-finite-depth-conj}}{cnot-finite-depth-conj.pdf}%
%width 3.2993in;  height 2.4794in;  depth 0pt;  original-width 6.5397in;
%original-height 4.9in;  cropleft "0";  croptop "1";  cropright "1";
%filename 'figures/CNOT-finite-depth-conj.pdf';file-properties "XNPEU";}}}%
%BeginExpansion
\begin{figure}
[ptb]
\begin{center}
\includegraphics[
natheight=4.900000in,
natwidth=6.539700in,
height=2.4794in,
width=3.2993in
]%
{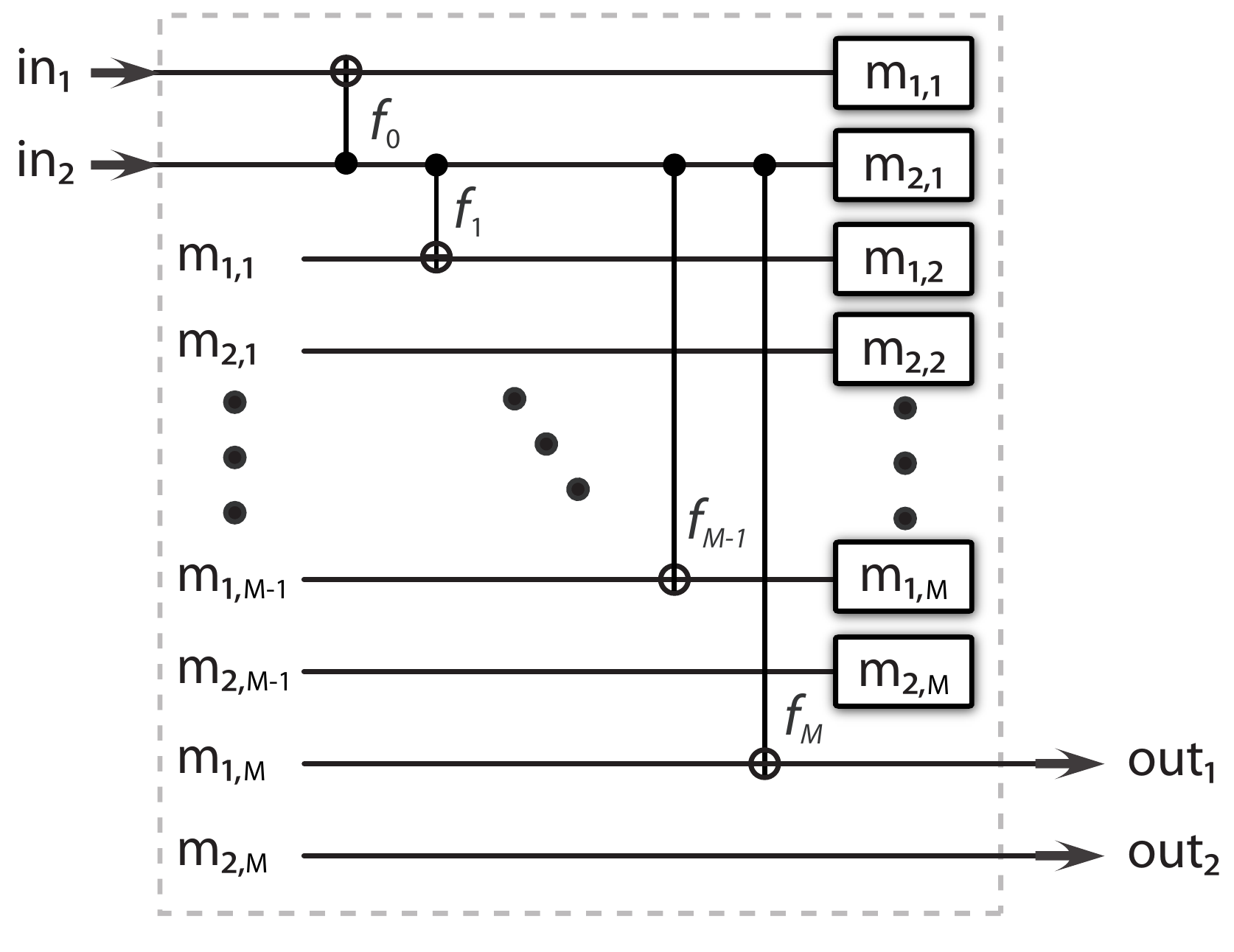}%
\caption{The circuit in the above figure implements the transformation in
(\ref{eq:finite-depth-poly-transform-conj}). Comparing the circuit in the
above figure to the one in Figure~\ref{fig:CNOT-finite-depth} reveals that it
merely flips the direction of the CNOT gates.}%
\label{fig:CNOT-finite-depth-conj}%
\end{center}
\end{figure}
%EndExpansion

\begin{theorem}
\label{thm:finite-depth-z}The circuit in
Figure~\ref{fig:CNOT-finite-depth-conj}\ implements the transformation in
(\ref{eq:finite-depth-poly-transform-conj}) and requires $M$ frames of memory qubits.
\end{theorem}

\begin{proof}
The proof follows analogously to the above proof by noting that the recursive
equations for the \textquotedblleft X\textquotedblright\ and \textquotedblleft
Z\textquotedblright\ variables interchange after reversing the direction of
the CNOT\ gates.
\end{proof}

\section{Memory Requirements for a CSS\ Quantum Convolutional Code}

\label{sec:memory-comp}Is there a general way for determining how much memory
a given code requires just by inspecting its stabilizer matrix? This section
answers this question with a theorem that determines the amount of memory that
a given CSS\ quantum convolutional code requires.

Ref.~\cite{ieee2007grassl} defines the individual constraint length, the
overall constraint length, and the memory of a quantum convolutional code in
analogy to the classical definitions \cite{book1999conv}. These definitions
are analogous to the classical definitions, but there does not seem to be an
operational interpretation of them in terms of the actual memory that a given
quantum convolutional code requires. We recall those definitions. The
constraint length $\nu_{i}$ for row $i$ of the stabilizer matrix is as
follows:%
\[
\nu_{i}\equiv\max_{j}\left\{  \max\left\{  \deg\left(  X_{ij}\left(  D\right)
,\deg\left(  Z_{ij}\left(  D\right)  \right)  \right)  \right\}  \right\}
\]
The overall constraint length $\nu$ is the sum of the individual constraint
lengths:%
\[
\nu\equiv\sum_{i}\nu_{i}.
\]
The memory $m$ is%
\[
m\equiv\max_{i}\nu_{i}.
\]

We now consider a general technique for computing the memory requirements of a
CSS\ quantum convolutional code, and the resulting formula does not correspond
to the above definitions. We exploit the Grassl-R\"{o}tteler algorithm for
encoding CSS\ codes \cite{ieee2007grassl}. This algorithm consists of a
sequence of Hadamards, a cascade of CNOTs, another sequence of Hadamards, and
another cascade of CNOTs. Here, I give a slightly simplfied algorithm that
does not require any Hadamard gates. Suppose that a quantum convolutional code
has the following stabilizer matrix:%
\begin{equation}
\left[  \left.
\begin{array}
[c]{c}%
H_{1}\left(  D\right)  \\
0
\end{array}
\right\vert
\begin{array}
[c]{c}%
0\\
H_{2}\left(  D\right)
\end{array}
\right]  .\label{eq:stab-matrix}%
\end{equation}
We can determine an encoding algorithm by looking at a series of steps to
decode the above quantum convolutional code. Assume that the matrices
$H_{1}\left(  D\right)  $ and $H_{2}\left(  D\right)  $ correspond to
noncatastrophic, delay-free check matrices so that they each have a Smith
normal form \cite{book1999conv,ieee2007grassl}:%
\[
H_{i}\left(  D\right)  =A_{i}\left(  D\right)  \left[
\begin{array}
[c]{cc}%
I & 0
\end{array}
\right]  B_{i}\left(  D\right)  ,
\]
where $i=1,2$. If the matrices $A_{i}\left(  D\right)  $ for $i=1,2$ are not
equal to the identity matrix, we can premultiply $H_{i}\left(  D\right)  $
with the inverse matrix $A_{i}^{-1}\left(  D\right)  $ for $i=1,2$. These row
operations do not affect the error-correcting properties of the quantum
convolutional code and give an equivalent code. Let us redefine the matrices
$H_{i}\left(  D\right)  $ as follows:%
\[
H_{i}\left(  D\right)  \equiv\left[
\begin{array}
[c]{cc}%
I & 0
\end{array}
\right]  B_{i}\left(  D\right)  ,
\]
for $i=1,2$. We can then write each matrix $B_{i}\left(  D\right)  $ as
follows:%
\[
B_{i}\left(  D\right)  =\left[
\begin{array}
[c]{c}%
H_{i}\left(  D\right)  \\
\tilde{H}_{i}\left(  D\right)
\end{array}
\right]  .
\]
Consider again the stabilizer matrix in (\ref{eq:stab-matrix}). Use
CNOT\ gates to perform the elementary column operations in the matrix
$B_{2}^{-1}\left(  D\right)  $. These operations postmultiply entries in the
\textquotedblleft X\textquotedblright\ matrix with the matrix $B_{2}%
^{-1}\left(  D\right)  $ and postmultiply entries in the \textquotedblleft
Z\textquotedblright\ matrix with the matrix $B_{2}^{T}\left(  D^{-1}\right)
$. The stabilizer matrix in (\ref{eq:stab-matrix}) transforms to the following
matrix:%
\[
\left[  \left.
\begin{array}
[c]{cc}%
H_{1}\left(  D\right)  H_{2}^{T}\left(  D^{-1}\right)   & H_{1}\left(
D\right)  \tilde{H}_{2}^{T}\left(  D^{-1}\right)  \\
0 & 0
\end{array}
\right\vert
\begin{array}
[c]{cc}%
0 & 0\\
I & 0
\end{array}
\right]  .
\]
The matrix $H_{1}\left(  D\right)  H_{2}^{T}\left(  D^{-1}\right)  $ is null
because the code is a CSS\ code and satisfies the dual-containing constraint.
The stabilizer matrix for the code is then as follows:%
\[
\left[  \left.
\begin{array}
[c]{cc}%
0 & H_{1}\left(  D\right)  \tilde{H}_{2}^{T}\left(  D^{-1}\right)  \\
0 & 0
\end{array}
\right\vert
\begin{array}
[c]{cc}%
0 & 0\\
I & 0
\end{array}
\right]  .
\]
Compute the Smith form of the matrix $H_{1}\left(  D\right)  \tilde{H}_{2}%
^{T}\left(  D^{-1}\right)  $:%
\[
H_{1}\left(  D\right)  \tilde{H}_{2}^{T}\left(  D^{-1}\right)  =A_{3}\left(
D\right)  \left[
\begin{array}
[c]{cc}%
I & 0
\end{array}
\right]  B_{3}\left(  D\right)  .
\]
Perform the row operations in $A_{3}^{-1}\left(  D\right)  $ on the first set
of rows. Finally, perform the conjugate CNOT\ gates corresponding to the
entries in $I\oplus B_{3}^{-1}\left(  D\right)  $---implying that we perform
them only on the last few qubits. These operations postmultiply the
\textquotedblleft X\textquotedblright\ matrix by the matrix $I\oplus B_{3}%
^{T}\left(  D^{-1}\right)  $ and postmultiply the \textquotedblleft
Z\textquotedblright\ matrix by the matrix $I\oplus B_{3}^{-1}\left(  D\right)
$. These operations then produce the following stabilizer matrix:%
\[
\left[  \left.
\begin{array}
[c]{ccc}%
0 & I & 0\\
0 & 0 & 0
\end{array}
\right\vert
\begin{array}
[c]{ccc}%
0 & 0 & 0\\
I & 0 & 0
\end{array}
\right]  .
\]
We are done at this point. These decoding operations give the following
transformations for the \textquotedblleft X\textquotedblright\ matrix:%
\[
B_{2}^{-1}\left(  D\right)  \left(  I\oplus B_{3}^{T}\left(  D^{-1}\right)
\right)  ,
\]
and the following transformations for the \textquotedblleft
Z\textquotedblright\ matrix:%
\[
B_{2}^{T}\left(  D^{-1}\right)  \left(  I\oplus B_{3}^{-1}\left(  D\right)
\right)  .
\]
To encode the quantum convolutional code, we perform the above operations in
the reverse order. For encoding, the following transformations postmultiply
the \textquotedblleft X\textquotedblright\ matrix:%
\[
E_{X}\left(  D\right)  \equiv\left(  I\oplus\left(  B_{3}^{T}\right)
^{-1}\left(  D^{-1}\right)  \right)  B_{2}\left(  D\right)  ,
\]
and the following transformations for the \textquotedblleft
Z\textquotedblright\ matrix:%
\[
E_{Z}\left(  D\right)  \equiv\left(  I\oplus B_{3}\left(  D\right)  \right)
\left(  B_{2}^{T}\right)  ^{-1}\left(  D^{-1}\right)  .
\]
The overall encoding matrix is
\[
B\left(  D\right)  \equiv\left[  \left.
\begin{array}
[c]{c}%
E_{Z}\left(  D\right)  \\
0
\end{array}
\right\vert
\begin{array}
[c]{c}%
0\\
E_{X}\left(  D\right)
\end{array}
\right]  .
\]
(See Ref.~\cite{ieee2007grassl} for a more detailed analysis of this algorithm).

I now give a theorem that determines the amount of memory that a CSS\ quantum
convolutional code requires.

\begin{theorem}
\label{thm:memory-CSS}The number $m$ of frames of memory qubits required for a
CSS\ quantum convolutional code encoded with the Grassl-R\"{o}tteler encoding
algorithm is upper bounded by the absolute degree of $B\left(  D\right)  $:%
\[
m\leq\left\vert \deg\right\vert \left(  B\left(  D\right)  \right)  .
\]

\end{theorem}

\begin{proof}
I employ an inductive method of proof. The above encoding algorithm for a
CSS\ quantum convolutional code demonstrates that we only have to consider how
CNOT\ gates combine together in a quantum shift register construction. We can
map each elementary CNOT\ operation to a quantum shift register circuit and
connect its outputs to the inputs of the quantum shift register circuit for
the next elementary CNOT\ operation. This technique is wasteful with respect
to memory, but the proof of this theorem shows all the ways that we can reduce
the amount of memory when combining quantum shift register circuits
corresponding to CNOT\ operations. The result of the theorem then gives a
simple formula for determining the amount of memory that a CSS\ quantum
convolutional code requires.

For the base step of the proof, consider that a CNOT\ gate from qubit $i$ to
qubit $j$ in a frame delayed by $l$ requires at most $l$ frames of memory
qubits. This result follows by extending the circuit of
Figure~\ref{fig:two-delay-CNOT}. The polynomial matrix for this CNOT\ gate
that acts on the $i^{\text{th}}$ and $j^{\text{th}}$ qubits is as follows:%
\[
\left[  \left.
\begin{array}
[c]{cc}%
1 & 0\\
D^{-l} & 1\\
0 & 0\\
0 & 0
\end{array}
\right\vert
\begin{array}
[c]{cc}%
0 & 0\\
0 & 0\\
1 & D^{l}\\
0 & 1
\end{array}
\right]  ,
\]
and has an absolute degree of $\left\vert l\right\vert $. We abbreviate the
above transformation as CNOT$\left(  i,j\right)  \left(  D^{l}\right)  $. So
the theorem holds for this base case.

Now consider two CNOT\ gates that have the same source qubits, but the source
of one of them acts on a target qubit in a frame delayed by $l_{0}$ and the
source of another acts on a target qubit in a frame delayed by $l_{1}$.
Suppose, without loss of generality, that $\left\vert l_{1}\right\vert
>\left\vert l_{0}\right\vert $ (these integers can be negative---we should use
the term \textquotedblleft advanced by\textquotedblright\ instead of
\textquotedblleft delayed by\textquotedblright\ for this case). This
combination is a special case of Theorems~\ref{thm:finite-depth-x} and
\ref{thm:finite-depth-z} and, therefore, we can implement this gate with
$\left\vert l_{1}\right\vert $ frames of memory qubits. The polynomial matrix
for the first CNOT\ is CNOT$\left(  i,j\right)  \left(  D^{l_{0}}\right)  $
and that for the second is CNOT$\left(  i,j\right)  \left(  D^{l_{1}}\right)
$. It is straightforward to check that the polynomial matrix for the combined
operation is CNOT$\left(  i,j\right)  \left(  D^{l_{0}}+D^{l_{1}}\right)  $.
The theorem thus holds for this case because the absolute degree of
CNOT$\left(  i,j\right)  \left(  D^{l_{0}}+D^{l_{1}}\right)  $ is $\left\vert
l_{1}\right\vert $.

The theorem similarly holds if two CNOT gates have the same source qubits but
have target qubits that do not have the same index within their given frame.
It also holds if two CNOT\ gates have the same target qubits but have source
qubits that do not have the same index within their given frame. The
polynomial matrices for the first case are CNOT$\left(  i,j\right)  \left(
D^{l_{0}}\right)  $ and CNOT$\left(  i,k\right)  \left(  D^{l_{1}}\right)  $
where WLOG $\left\vert l_{1}\right\vert >\left\vert l_{0}\right\vert $. These
two polynomial matrices commute. One can construct a quantum shift register
circuit with the techniques in this paper, and this circuit uses $\left\vert
l_{1}\right\vert $ frames of memory qubits. It is straightforward to check
that the absolute degree of the multiplication of matrices is $\left\vert
l_{1}\right\vert $. A similar symmetric analysis applies to the other case
where the target qubits are the same but the source qubits are different. The
main reason the theorem holds in these scenarios is that the polynomial matrix
representations of these gates commute with one another. Any time the
polynomial representations commute, the corresponding gates in the cascaded
quantum shift registers commute through memory so that the maximum amount of
frames of memory qubits is equal to the absolute degree of the entries in the
multiplication of the polynomial matrices.

Suppose the source qubits and target qubits of the two CNOT\ gates do not
intersect in any way. Then their polynomial matrix representations commute and
the amount of memory required is again equal to the absolute degree of the
polynomial matrices corresponding to the CNOT\ gates. An example is
CNOT$\left(  i,j\right)  \left(  D^{l_{1}}\right)  $ and CNOT\ $\left(
k,l\right)  \left(  D^{l_{0}}\right)  $ where $i\neq j\neq k\neq l$ and
WLOG\ $\left\vert l_{1}\right\vert >\left\vert l_{0}\right\vert $. One can use
the techniques in this paper to construct a combined quantum shift register
circuit that requires $\left\vert l_{1}\right\vert $ frames of memory qubits.

Suppose the index of source qubit of the first CNOT\ gate is the same as the
index of the target of the second CNOT\ gate, but the index of the target of
the first is different from the index of the source qubit of the second. An
example of this scenario is CNOT$\left(  i,j\right)  \left(  D^{l_{0}}\right)
$ followed by CNOT$\left(  k,i\right)  \left(  D^{l_{1}}\right)  $ where
$l_{0}$ and $l_{1}$ are any integers and $\left\vert l_{1}\right\vert
>\left\vert l_{0}\right\vert $ WLOG. The multiplication of the two polynomial
matrices gives the following polynomial matrix:%
\[
\left[  \left.
\begin{array}
[c]{ccc}%
1 & 0 & D^{-l_{1}}\\
D^{-l_{0}} & 1 & D^{-\left(  l_{1}+l_{0}\right)  }\\
0 & 0 & 1\\
0 & 0 & 0\\
0 & 0 & 0\\
0 & 0 & 0
\end{array}
\right\vert
\begin{array}
[c]{ccc}%
0 & 0 & 0\\
0 & 0 & 0\\
0 & 0 & 0\\
1 & D^{l_{0}} & 0\\
0 & 1 & 0\\
D^{l_{1}} & 0 & 1
\end{array}
\right]  ,
\]
where the indices $i$, $j$, and $k$ correspond to the first, second, and third
columns of the above \textquotedblleft Z\textquotedblright\ and
\textquotedblleft X\textquotedblright\ submatrices. It is again
straightforward using the technique in this paper to construct a quantum shift
register circuit that uses memory equal to the absolute degree of the above
polynomial matrix. The circuit uses $\left\vert l_{1}\right\vert $ frames of
memory qubits in the case that $l_{1}$ is positive and $l_{0}$ is negative and
vice versa and uses $\left\vert l_{1}+l_{0}\right\vert $ frames of memory
qubits in the case that $l_{0}$ and $l_{1}$ are both positive or both negative.

The last scenario to consider is when the index of the source qubit of the
first CNOT\ gate is the same as the index of the target of the second
CNOT\ gate, and the index of the target of the first CNOT\ gate is the same as
the index of the source of the second CNOT\ gate. An example of this scenario
is CNOT$\left(  i,j\right)  \left(  D^{l_{0}}\right)  $ followed by
CNOT$\left(  j,i\right)  \left(  D^{l_{1}}\right)  $, where $l_{0}$ and
$l_{1}$ are any integers and $\left\vert l_{1}\right\vert >\left\vert
l_{0}\right\vert $ WLOG. The multiplication of the two polynomial matrices
gives the following polynomial matrix:%
\[
\left[  \left.
\begin{array}
[c]{cc}%
1 & D^{-l_{1}}\\
D^{-l_{0}} & 1+D^{-\left(  l_{0}+l_{1}\right)  }\\
0 & 0\\
0 & 0
\end{array}
\right\vert
\begin{array}
[c]{cc}%
0 & 0\\
0 & 0\\
1+D^{l_{0}+l_{1}} & D^{l_{0}}\\
D^{l_{1}} & 1
\end{array}
\right]  ,
\]
where the indices $i$ and $j$ correspond to the first and second columns of
the above \textquotedblleft Z\textquotedblright\ and \textquotedblleft
X\textquotedblright\ submatrices. It is again straightforward to construct a
quantum shift register using the techniques in this paper that uses a number
of frames of memory qubits equal to the absolute degree of the polynomial matrix.

The inductive step follows by considering that any arbitrary encoding with
CNOTs is a sequence of elementary column operations of the form:%
\[
B\left(  D\right)  =B_{\left(  1\right)  }\left(  D\right)  B_{\left(
2\right)  }\left(  D\right)  \cdots B_{\left(  p\right)  }\left(  D\right)  ,
\]
where $p$ is the total number of elementary operations and the above
decomposition is a \textit{particular} decomposition of the matrix $B\left(
D\right)  $ into elementary operations. Suppose the above encoding matrix
requires $m$ frames of memory qubits and $m$ is also the absolute degree of
$B\left(  D\right)  $. Suppose we cascade another elementary encoding
operation with matrix representation $B_{\left(  p+1\right)  }\left(
D\right)  $. If $B_{\left(  p+1\right)  }\left(  D\right)  $ commutes with
$B_{\left(  p\right)  }\left(  D\right)  $, then it increases the absolute
degree of the resulting matrix $B\left(  D\right)  $ and the memory required
for the quantum shift register circuit only if it has a higher absolute degree
than $B_{\left(  p\right)  }\left(  D\right)  $. The case is thus reducible to
the case where it does not commute with $B_{\left(  p\right)  }\left(
D\right)  $. So, suppose $B_{\left(  p+1\right)  }\left(  D\right)  $ does not
commute with $B_{\left(  p\right)  }\left(  D\right)  $. There are two ways in
which this non-commutativity can happen and I detailed them above. The
analysis above for both cases shows that the absolute degree and the number of
frames of memory qubits increase by the same amount depending whether $l_{0}$
and $l_{1}$ are positive or negative.
\end{proof}

\begin{corollary}
A Type\ I\ CSS\ entanglement-assisted quantum convolutional code
\cite{arx2007wildeEAQCC}\ encoded with the Grassl-R\"{o}tteler encoding
algorithm requires $m$ frames of memory qubits, where%
\[
m\leq\left\vert \deg\right\vert \left(  B\left(  D\right)  \right)  .
\]
The matrix $B\left(  D\right)  $ is the polynomial matrix representation of
the encoding operations of the entanglement-assisted code.
\end{corollary}

\begin{proof}
A Type I CSS\ entanglement-assisted convolutional code is one that has a
finite-depth encoding and decoding circuit. It is possible to show that the
encoding circuit consists entirely of CNOT\ gates. The proof proceeds
analogously to the proof of the above theorem.
\end{proof}

\section{Other Operations in the Finite-Depth Shift-Invariant Clifford Group}

\label{sec:finite-depth-Clifford}CNOT gates are not the only gates that are
useful for encoding a quantum convolutional code. The Hadamard gate, the Phase
gate, and the controlled-Phase gate are also useful and are in the
finite-depth shift-invariant Clifford group \cite{isit2006grassl}.

There is no need to formulate a primitive quantum shift register circuit for
the Hadamard gate or the Phase gate---the implementation is trivial and does
not require memory qubits.

The controlled-phase gate is useful for implementation with a quantum shift
register circuit because it acts on two qubits. There are two types of a
quantum shift register circuit that we can develop with a controlled-Phase
gate. The first type is similar to that for the finite-depth CNOT quantum
shift register circuit because it involves two qubits per frame. The second
type is different because it involves only one qubit per frame.

\subsection{Finite-Depth Controlled-Phase Gate with Two Qubits per Frame}

Suppose that we have two qubits on which we would like to perform a
finite-depth controlled-Phase gate operation. The Pauli operators for these
qubits are as follows:%
\[
\left[  \left.
\begin{array}
[c]{cc}%
1 & 0\\
0 & 1\\
0 & 0\\
0 & 0
\end{array}
\right\vert
\begin{array}
[c]{cc}%
0 & 0\\
0 & 0\\
1 & 0\\
0 & 1
\end{array}
\right]  .
\]
A general shift-invariant finite-depth controlled-Phase gate\ operation
translates the above set of operators to the following set:%
\begin{equation}
\left[  \left.
\begin{array}
[c]{cc}%
1 & 0\\
0 & 1\\
0 & f\left(  D\right) \\
f\left(  D^{-1}\right)  & 0
\end{array}
\right\vert
\begin{array}
[c]{cc}%
0 & 0\\
0 & 0\\
1 & 0\\
0 & 1
\end{array}
\right]  , \label{eq:finite-depth-controlled-phase}%
\end{equation}
where $f\left(  D\right)  $ is some arbitrary binary polynomial:%
\[
f\left(  D\right)  =\sum_{i=0}^{M}f_{i}D^{i}.
\]

\begin{theorem}
\label{thm:finite-depth-controlled-Phase}The circuit in
Figure~\ref{fig:cphase-finite-depth}\ implements the transformation in
(\ref{eq:finite-depth-controlled-phase}), and it requires no more than $M$
frames of memory qubits.
\end{theorem}

\begin{proof}
The proof of this theorem is similar to that of the previous theorems. We can
formulate a system of recursive equations involving binary variables. Let us
label the bit representations of the $X$ Pauli operators for all the qubits as
follows:%
\[
x_{1}^{\prime},x_{2}^{\prime},m_{1,1}^{x},m_{2,1}^{x},m_{1,2}^{x},m_{2,2}%
^{x},\ldots,m_{1,M}^{x},m_{2,M}^{x},x_{1},x_{2},
\]
where the primed variables are the outputs and the unprimed are the inputs.
Let us label the bit representations of the $Z$ Pauli operators similarly:%
\[
z_{1}^{\prime},z_{2}^{\prime},m_{1,1}^{z},m_{2,1}^{z},m_{1,2}^{z},m_{2,2}%
^{z},\ldots,m_{1,M}^{z},m_{2,M}^{z},z_{1},z_{2}.
\]%
%TCIMACRO{\FRAME{ftbpFU}{3.039in}{2.284in}{0pt}{\Qcb{The circuit in the above
%figure implements the transformation in
%(\ref{eq:finite-depth-controlled-phase}).}}{\Qlb{fig:cphase-finite-depth}%
%}{cphase-finite-depth.pdf}{\special{ language "Scientific Word";
%type "GRAPHIC";  maintain-aspect-ratio TRUE;  display "USEDEF";
%valid_file "F";  width 3.039in;  height 2.284in;  depth 0pt;
%original-width 6.5397in;  original-height 4.9in;  cropleft "0";  croptop "1";
%cropright "1";  cropbottom "0";
%filename 'figures/CPhase-finite-depth.pdf';file-properties "XNPEU";}}}%
%BeginExpansion
\begin{figure}
[ptb]
\begin{center}
\includegraphics[
natheight=4.900000in,
natwidth=6.539700in,
height=2.284in,
width=3.039in
]%
{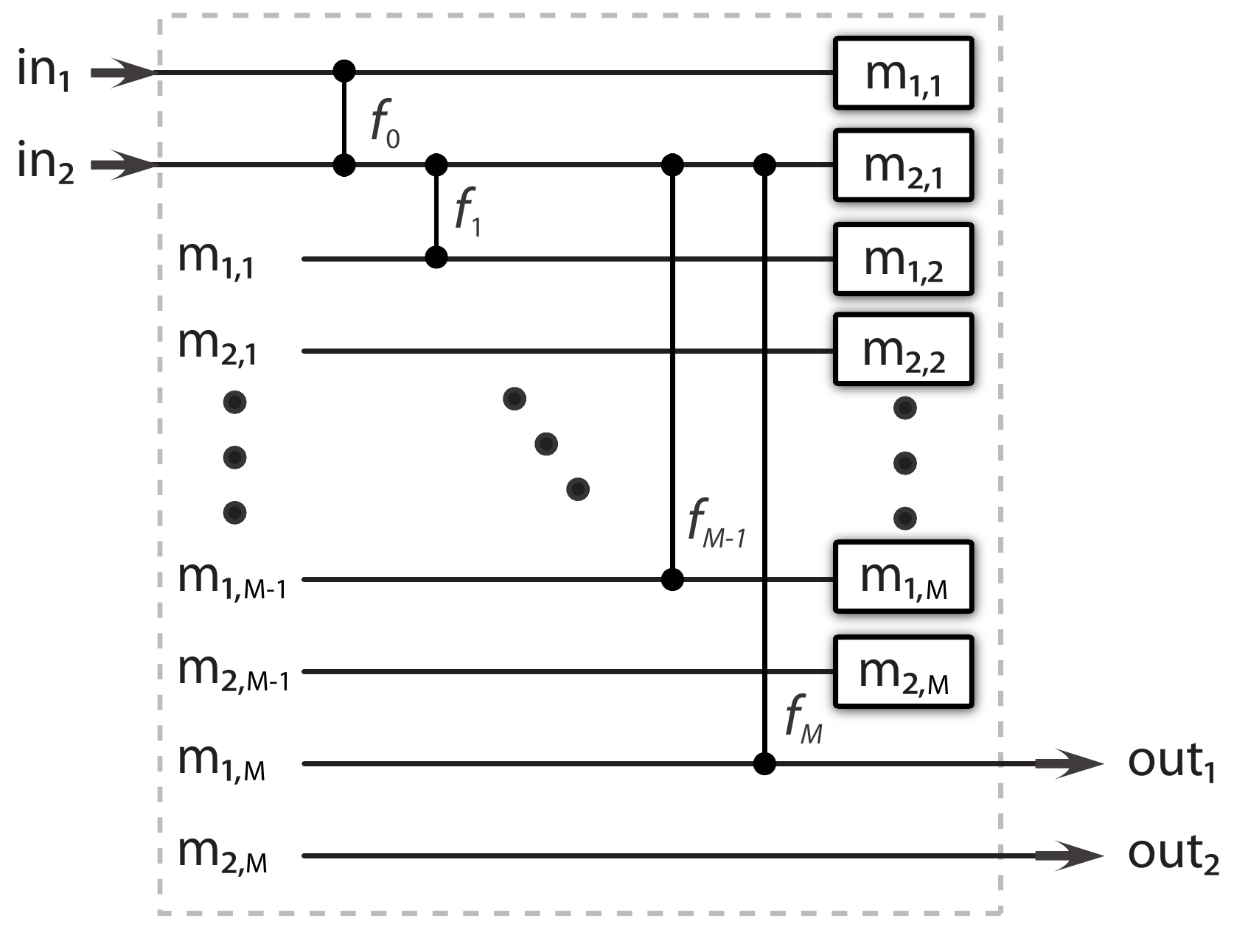}%
\caption{The circuit in the above figure implements the transformation in
(\ref{eq:finite-depth-controlled-phase}).}%
\label{fig:cphase-finite-depth}%
\end{center}
\end{figure}
%EndExpansion
The circuit in Figure~\ref{fig:cphase-finite-depth} implements the following
set of recursive \textquotedblleft X\textquotedblright\ equations:%
\begin{align*}
x_{1}^{\prime}\left[  n\right]   &  =m_{1,M}^{x}\left[  n-1\right]  ,\\
x_{2}^{\prime}\left[  n\right]   &  =m_{2,M}^{x}\left[  n-1\right]  ,\\
m_{1,1}^{x}\left[  n\right]   &  =x_{1}\left[  n\right]  ,\\
m_{2,1}^{x}\left[  n\right]   &  =x_{2}\left[  n\right]  ,
\end{align*}
and $\forall i=2\ldots M$,%
\begin{align*}
m_{1,i}^{x}\left[  n\right]   &  =m_{1,i-1}^{x}\left[  n-1\right]  ,\\
m_{2,i}^{x}\left[  n\right]   &  =m_{2,i-1}^{x}\left[  n-1\right]  .
\end{align*}
The set of \textquotedblleft Z\textquotedblright\ recursive equations is as
follows:%
\begin{align*}
z_{1}^{\prime}\left[  n\right]   &  =m_{1,M}^{z}\left[  n-1\right]
+f_{M}\ x_{2}\left[  n\right]  ,\\
z_{2}^{\prime}\left[  n\right]   &  =m_{2,M}^{z}\left[  n-1\right]  ,\\
m_{1,1}^{z}\left[  n\right]   &  =z_{1}\left[  n\right]  +f_{0}\ x_{2}\left[
n\right]  ,\\
m_{2,1}^{z}\left[  n\right]   &  =z_{2}\left[  n\right]  +f_{0}x_{1}\left[
n\right]  +\sum_{i=1}^{M}f_{i}m_{1,i}^{x}\left[  n-1\right]  ,
\end{align*}
and $\forall i=2,\ldots,M$,%
\begin{align*}
m_{1,i}^{z}\left[  n\right]   &  =m_{1,i-1}^{z}\left[  n-1\right]
+f_{i-1}\ x_{2}\left[  n\right]  ,\\
m_{2,i}^{z}\left[  n\right]   &  =m_{2,i-1}^{z}\left[  n-1\right]  .
\end{align*}
Simplifying\ the \textquotedblleft X\textquotedblright\ equations gives the
following two equations:%
\begin{align*}
x_{1}^{\prime}\left[  n\right]   &  =x_{1}\left[  n-M\right]  ,\\
x_{2}^{\prime}\left[  n\right]   &  =x_{2}\left[  n-M\right]  .
\end{align*}
Simplifying\ the \textquotedblleft Z\textquotedblright\ equations gives the
following two equations:%
\begin{align*}
z_{1}^{\prime}\left[  n\right]   &  =z_{1}\left[  n-M\right]  +\sum_{i=0}%
^{M}f_{i}x_{2}\left[  n-M+i\right]  ,\\
z_{2}^{\prime}\left[  n\right]   &  =z_{2}\left[  n-M\right]  +\sum_{i=0}%
^{M}f_{i}x_{1}\left[  n-M-i\right]  .
\end{align*}
Applying the $D$-transform to the above gives the following set of equations:%
\begin{align*}
x_{1}^{\prime}\left(  D\right)   &  =D^{M}x_{1}\left(  D\right)  ,\\
x_{2}^{\prime}\left(  D\right)   &  =D^{M}x_{2}\left(  D\right)  ,\\
z_{1}^{\prime}\left(  D\right)   &  =D^{M}\left(  z_{1}\left(  D\right)
+\sum_{i=0}^{M}f_{i}D^{-i}x_{2}\left(  D\right)  \right) \\
&  =D^{M}\left(  z_{1}\left(  D\right)  +f\left(  D^{-1}\right)  z_{2}\left(
D\right)  \right)  ,\\
z_{2}^{\prime}\left(  D\right)   &  =D^{M}z_{2}\left(  D\right)  +\sum
_{i=0}^{M}f_{i}D^{i}x_{1}\left(  D\right) \\
&  =D^{M}\left(  z_{2}\left(  D\right)  +f\left(  D\right)  x_{1}\left(
D\right)  \right)  ,
\end{align*}
Rewriting the above set of equations as a matrix transformation reveals that
it is equivalent to the transformation in
(\ref{eq:finite-depth-controlled-phase}):%
\[
\left[  \left.
\begin{array}
[c]{cc}%
1 & 0\\
0 & 1\\
0 & f\left(  D\right) \\
f\left(  D^{-1}\right)  & 0
\end{array}
\right\vert
\begin{array}
[c]{cc}%
0 & 0\\
0 & 0\\
1 & 0\\
0 & 1
\end{array}
\right]  D^{M}.
\]
Postmultiplying the following vector by the above transformation%
\[
\left[  \left.
\begin{array}
[c]{cc}%
z_{1}\left(  D\right)  & z_{2}\left(  D\right)
\end{array}
\right\vert
\begin{array}
[c]{cc}%
x_{1}\left(  D\right)  & x_{2}\left(  D\right)
\end{array}
\right]  ,
\]
gives the following output vector:%
\[
\left[  \left.
\begin{array}
[c]{cc}%
z_{1}^{\prime}\left(  D\right)  & z_{2}^{\prime}\left(  D\right)
\end{array}
\right\vert
\begin{array}
[c]{cc}%
x_{1}^{\prime}\left(  D\right)  & x_{2}^{\prime}\left(  D\right)
\end{array}
\right]  .
\]

\end{proof}

\subsection{Finite-Depth Controlled-Phase Gate with One Qubit per Frame}

Suppose that we have one qubit on which we would like to perform a
finite-depth controlled-Phase gate operation. The Pauli operators for this
qubit are as follows:%
\[
\left[  \left.
\begin{array}
[c]{c}%
1\\
0
\end{array}
\right\vert
\begin{array}
[c]{c}%
0\\
1
\end{array}
\right]  .
\]
A general shift-invariant finite-depth controlled-Phase gate\ operation
translates the above set of operators to the following set:%
\begin{equation}
\left[  \left.
\begin{array}
[c]{c}%
1\\
f\left(  D\right)  +f\left(  D^{-1}\right)
\end{array}
\right\vert
\begin{array}
[c]{c}%
0\\
1
\end{array}
\right]  , \label{eq:finite-depth-controlled-phase-single}%
\end{equation}
where $f\left(  D\right)  $ is some arbitrary binary polynomial:%
\[
f\left(  D\right)  =\sum_{i=1}^{M}f_{i}D^{i}.
\]

\begin{theorem}
\label{thm:finite-depth-controlled-Phase-single}The circuit in
Figure~\ref{fig:cphase-finite-depth-single}\ implements the transformation in
(\ref{eq:finite-depth-controlled-phase-single}) and it requires $M$ frames of
memory qubits.
\end{theorem}

\begin{proof}
The proof of this theorem is similar to that of the previous theorems. We can
formulate a system of recursive equations involving binary variables. Let us
label the bit representations of the $X$ Pauli operators for all the qubits as
follows:%
\[
x^{\prime},m_{1}^{x},m_{2}^{x},\ldots,m_{M}^{x},x,
\]
where the primed variables are the outputs and the unprimed are the inputs.
Let us label the bit representations of the $Z$ Pauli operators similarly:%
\[
z^{\prime},m_{1}^{z},m_{2}^{z},\ldots,m_{M}^{z},z.
\]%
%TCIMACRO{\FRAME{ftbpFU}{3.039in}{1.6544in}{0pt}{\Qcb{The circuit in the above
%figure implements the transformation in (\ref{fig:cphase-finite-depth-single}%
%).}}{\Qlb{fig:cphase-finite-depth-single}}{cphase-finite-depth-single.pdf}%
%width 3.039in;  height 1.6544in;  depth 0pt;  original-width 6.3529in;
%original-height 3.4333in;  cropleft "0";  croptop "1";  cropright "1";
%filename 'figures/CPhase-finite-depth-single.pdf';file-properties "XNPEU";}}}%
%BeginExpansion
\begin{figure}
[ptb]
\begin{center}
\includegraphics[
natheight=3.433300in,
natwidth=6.352900in,
height=1.6544in,
width=3.039in
]%
{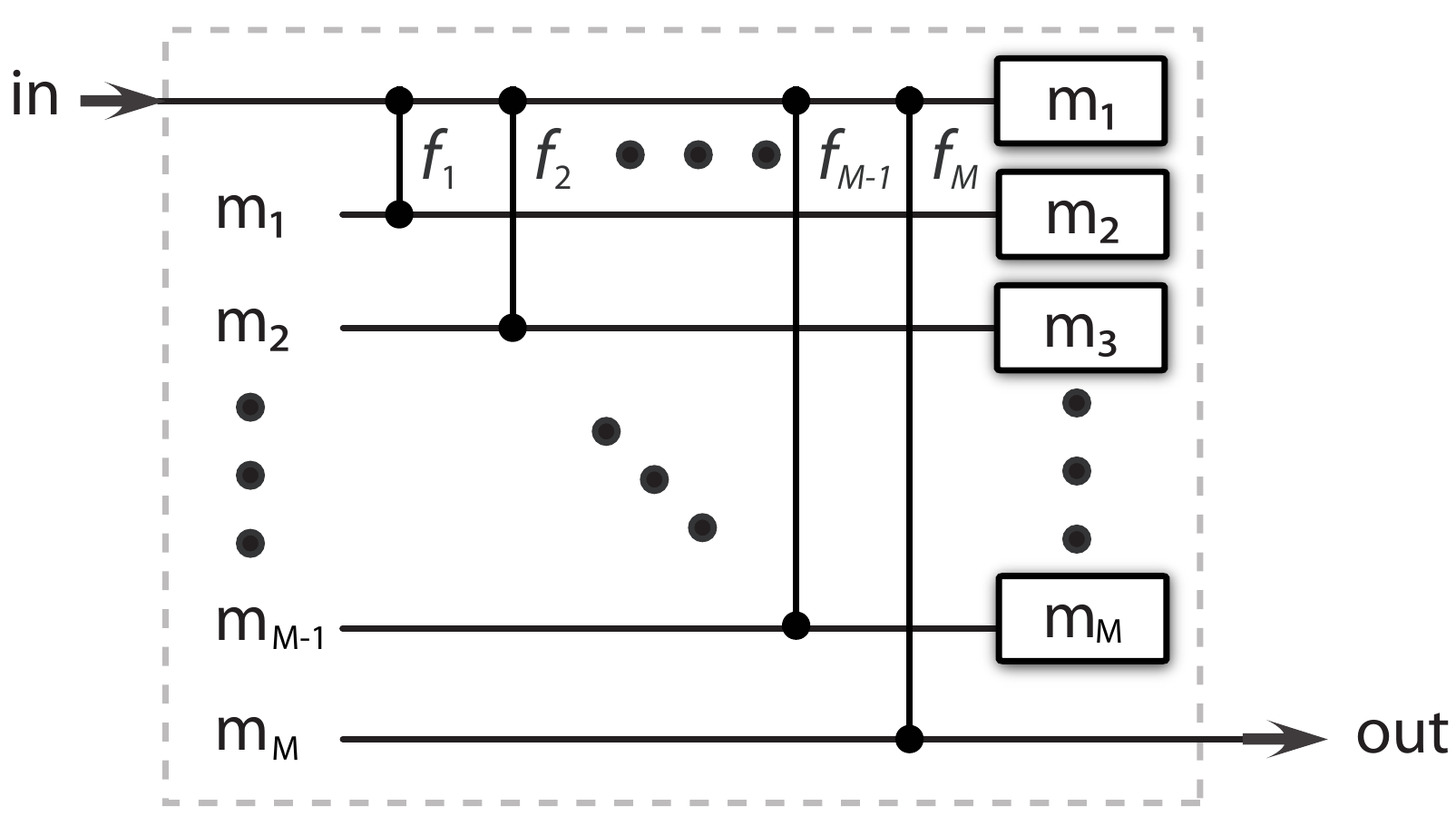}%
\caption{The circuit in the above figure implements the transformation in
(\ref{fig:cphase-finite-depth-single}).}%
\label{fig:cphase-finite-depth-single}%
\end{center}
\end{figure}
%EndExpansion
The circuit in Figure~\ref{fig:cphase-finite-depth-single} implements the
following set of recursive \textquotedblleft X\textquotedblright\ equations:%
\begin{align*}
x_{1}^{\prime}\left[  n\right]   &  =m_{M}^{x}\left[  n-1\right]  ,\\
m_{1}^{x}\left[  n\right]   &  =x\left[  n\right]  ,
\end{align*}
and $\forall i=2\ldots M$,%
\[
m_{i}^{x}\left[  n\right]  =m_{i-1}^{x}\left[  n-1\right]  .
\]
The set of \textquotedblleft Z\textquotedblright\ recursive equations is as
follows:%
\begin{align*}
z^{\prime}\left[  n\right]   &  =m_{M}^{z}\left[  n-1\right]  +f_{M}\ x\left[
n\right]  ,\\
m_{1}^{z}\left[  n\right]   &  =z\left[  n\right]  +\sum_{i=1}^{M}f_{i}%
m_{i}^{x}\left[  n-1\right]  ,
\end{align*}
and $\forall i=2,\ldots,M$,%
\[
m_{i}^{z}\left[  n\right]  =m_{i-1}^{z}\left[  n-1\right]  +f_{i-1}\ x\left[
n\right]  .
\]
Simplifying\ the \textquotedblleft X\textquotedblright\ equations gives the
following equation:%
\[
x^{\prime}\left[  n\right]  =x\left[  n-M\right]  .
\]
Simplifying\ the \textquotedblleft Z\textquotedblright\ equations gives the
following equation:%
\begin{multline*}
z^{\prime}\left[  n\right]  =z\left[  n-M\right]  +\sum_{i=1}^{M}f_{i}x\left[
n-M+i\right] \\
+\sum_{i=1}^{M}f_{i}x\left[  n-M-i\right]  .
\end{multline*}
Applying the $D$-transform to the above gives the following set of equations:%
\begin{align*}
x^{\prime}\left(  D\right)   &  =D^{M}x\left(  D\right)  ,\\
z^{\prime}\left(  D\right)   &  =D^{M}\left(  z\left(  D\right)  +\sum
_{i=1}^{M}f_{i}D^{-i}x\left(  D\right)  +\sum_{i=1}^{M}f_{i}D^{i}x\left(
D\right)  \right) \\
&  =D^{M}\left(  z\left(  D\right)  +\left(  f\left(  D^{-1}\right)  +f\left(
D\right)  \right)  x\left(  D\right)  \right)  ,
\end{align*}
Rewriting the above set of equations as a matrix transformation reveals that
it is equivalent to the transformation in
(\ref{eq:finite-depth-controlled-phase}):%
\[
\left[  \left.
\begin{array}
[c]{c}%
1\\
f\left(  D^{-1}\right)  +f\left(  D\right)
\end{array}
\right\vert
\begin{array}
[c]{c}%
0\\
1
\end{array}
\right]  D^{M}.
\]
Postmultiplying the following vector by the above transformation%
\[
\left[  \left.
\begin{array}
[c]{c}%
z\left(  D\right)
\end{array}
\right\vert
\begin{array}
[c]{c}%
x\left(  D\right)
\end{array}
\right]  ,
\]
gives the following output vector:%
\[
\left[  \left.
\begin{array}
[c]{c}%
z^{\prime}\left(  D\right)
\end{array}
\right\vert
\begin{array}
[c]{c}%
x^{\prime}\left(  D\right)
\end{array}
\right]  .
\]

\end{proof}

\section{Quantum Shift Register Encoding Circuit for the Forney-Grassl-Guha
Code}

\label{sec:forney-guha-grassl}
\begin{figure}
[ptb]
\begin{center}
\includegraphics[
natheight=9.072800in,
natwidth=9.460200in,
height=3.2007in,
width=3.2396in
]%
{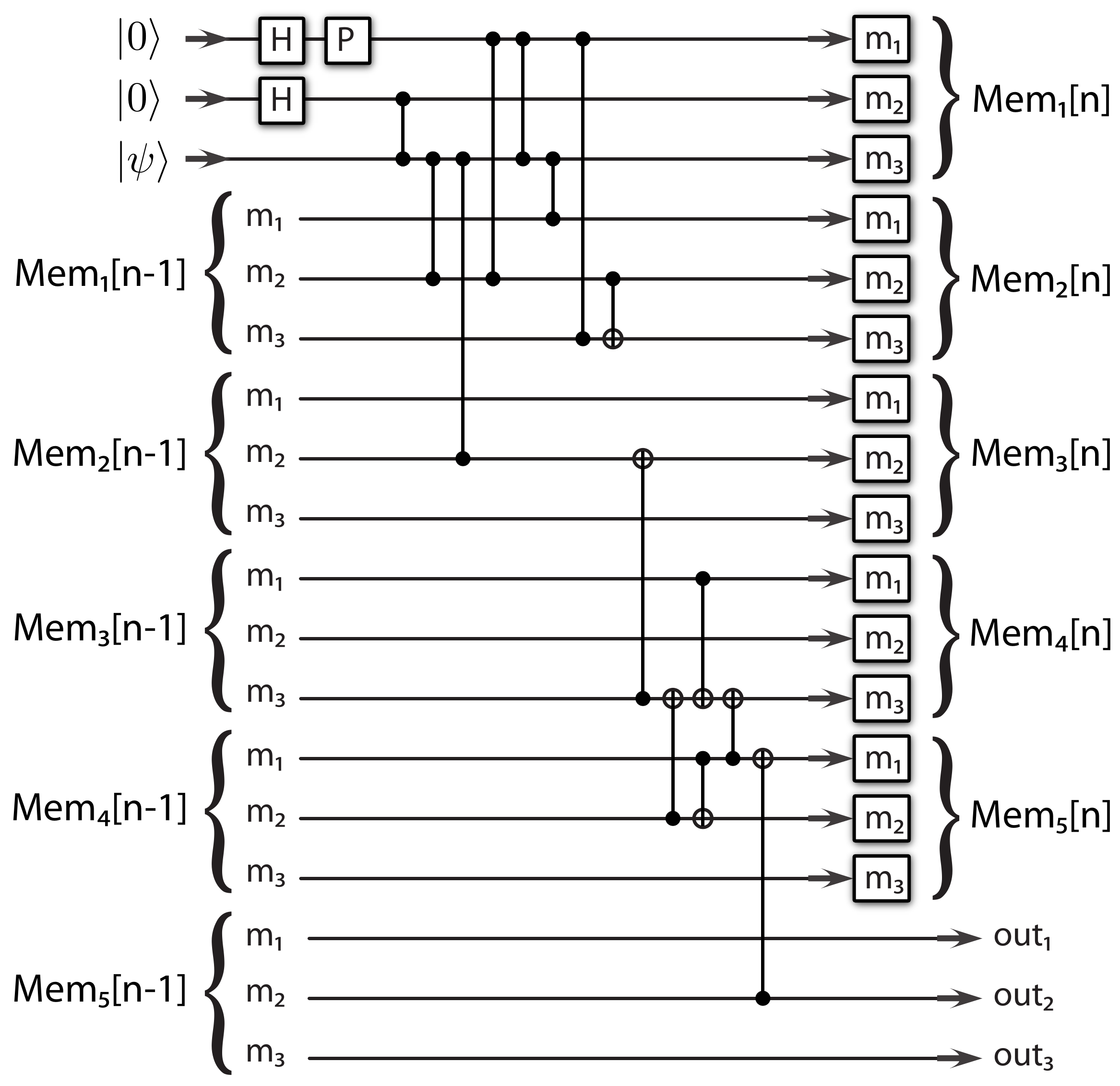}%
\caption{The above circuit encodes the Forney-Grassl-Guha code from
Ref.~\cite{ieee2007forney}.}%
\label{fig:FGG-circuit}%
\end{center}
\end{figure}

I now present another example of a quantum shift
register encoding circuit for a quantum convolutional code. The code that I
choose is the Forney-Grassl-Guha code from Section IIIB of
Ref.~\cite{ieee2007forney}. The code has three qubits per frame, and its
stabilizer matrix is%
\[
\left[  \left.
\begin{array}
[c]{ccc}%
1+D & 1 & 1+D\\
0 & D & D
\end{array}
\right\vert
\begin{array}
[c]{ccc}%
0 & D & D\\
1+D & 1+D & 1
\end{array}
\right]  .
\]
We can again employ the Grassl-R\"{o}tteler encoding algorithm
\cite{ieee2006grassl}\ to determine a sequence of encoding operations for this
code. This sequence of encoding operations is%
\begin{align*}
&  H\left(  1\right)  H\left(  2\right)  P\left(  1\right)  \text{C-PHASE}%
\left(  1,3\right)  \left(  D^{-1}+1+D\right) \\
&  \text{C-PHASE}\left(  1,2\right)  \left(  D^{-1}\right)  \text{C-PHASE}%
\left(  2,3\right)  \left(  1+D+D^{2}\right) \\
&  \text{CNOT}\left(  2,3\right)  \left(  1\right)  \text{CNOT}\left(
3,2\right)  \left(  D\right)  \text{CNOT}\left(  2,3\right)  \left(  D\right)
\\
&  \text{CNOT}\left(  1,2\right)  \left(  1\right)  \text{CNOT}\left(
1,3\right)  \left(  1+D\right)  \text{CNOT}\left(  2,1\right)  \left(
D\right)  ,
\end{align*}
where the order of operations goes from left to right and top to bottom,
$H\left(  i\right)  $ is a Hadamard gate on qubit $i$, and $P\left(  i\right)
$ is a Phase gate on qubit $i$. I use the technique in this paper to cascade
several quantum shift register circuits and commute gates through memory.
Figure~\ref{fig:FGG-circuit}\ depicts the quantum shift register circuit that
encodes the Forney-Grassl-Guha code.%
%TCIMACRO{\FRAME{ftbpFU}{3.2396in}{3.2007in}{0pt}{\Qcb{The above circuit
%encodes the Forney-Grassl-Guha code from Ref.~\cite{ieee2007forney}.}%
%}{\Qlb{fig:FGG-circuit}}{fgg-circuit.pdf}%
%width 3.2396in;  height 3.2007in;  depth 0pt;  original-width 9.4602in;
%original-height 9.0728in;  cropleft "0";  croptop "1";  cropright "1";
%cropbottom "0";  filename 'figures/FGG-circuit.pdf';file-properties "XNPEU";}%
%}}%
%BeginExpansion
%EndExpansion

\section{General Infinite-Depth Operations}

\label{sec:infinite-depth}We now turn to infinite-depth operations. Briefly,
infinite-depth operations can take a finite-weight Pauli operator to an
infinite-weight Pauli operator (similar to the way that an infinite-impulse
response filter can have an infinite-duration response to a finite-duration
input). Section~VI of Ref.~\cite{arx2007wildeEAQCC}\ discusses infinite-depth
Clifford operations. Here, I give a simplification of that discussion by
showing how to implement an arbitrary infinite-depth operation using quantum
shift register circuits.

Let $f\left(  D\right)  $ be some binary polynomial:%
\begin{equation}
f\left(  D\right)  =\sum_{i=0}^{M}f_{i}D^{i}. \label{eq:inf-depth-poly}%
\end{equation}
Suppose that we have one qubit on which we would like to perform an
infinite-depth controlled-NOT\ operation. The Pauli operators for this qubit
are as follows:%
\begin{equation}
\left[  \left.
\begin{array}
[c]{c}%
1\\
0
\end{array}
\right\vert
\begin{array}
[c]{c}%
0\\
1
\end{array}
\right]  . \label{eq:logical-operators}%
\end{equation}
A \textquotedblleft Z\textquotedblright\ infinite-depth operation transforms
the logical operators to be as follows:%
\begin{equation}
\left[  \left.
\begin{array}
[c]{c}%
1/f\left(  D^{-1}\right) \\
0
\end{array}
\right\vert
\begin{array}
[c]{c}%
0\\
f\left(  D\right)
\end{array}
\right]  , \label{eq:z-inf-depth}%
\end{equation}
where $1/f\left(  D^{-1}\right)  =D^{M}/D^{M}f\left(  D^{-1}\right)  $.

\begin{theorem}
\label{thm:z-inf-depth}The circuit in Figure~\ref{fig:z-inf-depth}\ implements
the transformation in (\ref{eq:z-inf-depth}) and requires $M$ memory qubits.%
%TCIMACRO{\FRAME{ftbpFU}{3.205in}{1.7452in}{0pt}{\Qcb{The quantum shift
%register circuit in the above figure implements the transformation in
%(\ref{eq:z-inf-depth}).}}{\Qlb{fig:z-inf-depth}}{z-inf-depth.pdf}%
%width 3.205in;  height 1.7452in;  depth 0pt;  original-width 6.3529in;
%original-height 3.4333in;  cropleft "0";  croptop "1";  cropright "1";
%cropbottom "0";  filename 'figures/z-inf-depth.pdf';file-properties "XNPEU";}%
%}}%
%BeginExpansion
\begin{figure}
[ptb]
\begin{center}
\includegraphics[
natheight=3.433300in,
natwidth=6.352900in,
height=1.7452in,
width=3.205in
]%
{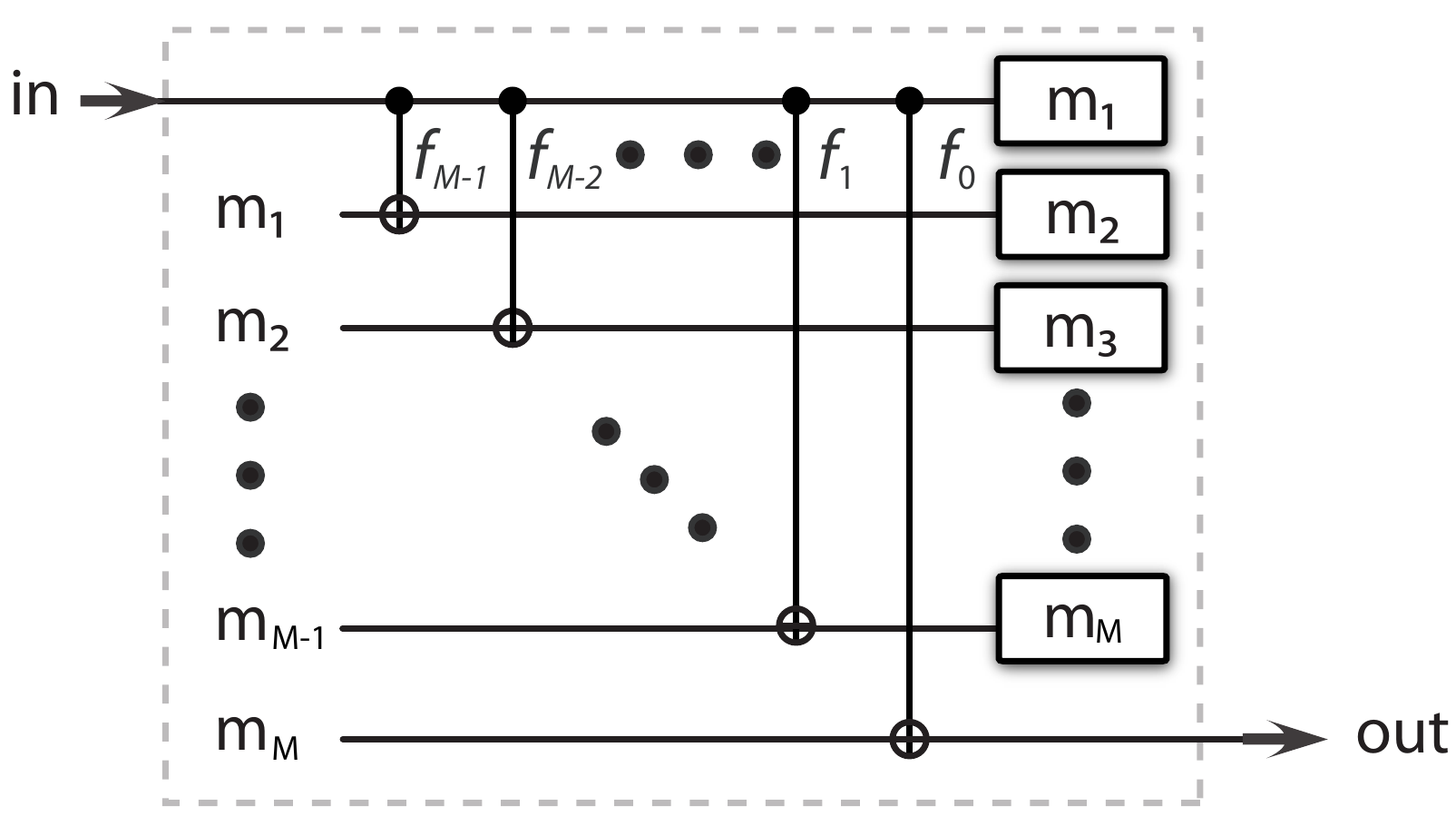}%
\caption{The quantum shift register circuit in the above figure implements the
transformation in (\ref{eq:z-inf-depth}).}%
\label{fig:z-inf-depth}%
\end{center}
\end{figure}
%EndExpansion

\end{theorem}

\begin{proof}
I use a similar linear system theoretic technique that exploits recursive
equations and the $D$-transform and assume without loss of generality that the
coefficient $f_{M}=1$. We use similar notation as before for the
\textquotedblleft X\textquotedblright\ and \textquotedblleft
Z\textquotedblright\ variables. We get the following set of \textquotedblleft
X\textquotedblright\ recursive equations:%
\begin{align*}
x^{\prime}\left[  n\right]   &  =m_{M}^{x}\left[  n-1\right]  +f_{0}x\left[
n\right]  ,\\
m_{1}^{x}\left[  n\right]   &  =x\left[  n\right]  ,
\end{align*}
and $\forall i=2,\ldots,M$,%
\[
m_{i}^{x}\left[  n\right]  =m_{i-1}^{x}\left[  n-1\right]  +f_{M-i+1}x\left[
n\right]  .
\]
The \textquotedblleft Z\textquotedblright\ recursive equations are as follows:%
\begin{align*}
z^{\prime}\left[  n\right]   &  =m_{M}^{z}\left[  n-1\right]  ,\\
m_{1}^{z}\left[  n\right]   &  =z\left[  n\right]  +\sum_{i=0}^{M-1}%
f_{i}m_{M-i}^{z}\left[  n-1\right]  ,
\end{align*}
and $\forall i=2\ldots M$,%
\[
m_{i}^{z}\left[  n\right]  =m_{i-1}^{z}\left[  n-1\right]  .
\]
The first set of \textquotedblleft X\textquotedblright\ recursive equations
reduces to the following equation by substitution:%
\begin{equation}
x^{\prime}\left[  n\right]  =\sum_{i=0}^{M}f_{i}x\left[  n-i\right]  .
\label{eq:inf-depth-rec-simple}%
\end{equation}
We can reduce the \textquotedblleft Z\textquotedblright\ equations by first
noticing that we can rewrite the equation for $m_{1}^{z}$ as follows:%
\[
m_{1}^{z}\left[  n\right]  +f_{M-1}m_{1}^{z}\left[  n-1\right]  =z\left[
n\right]  +\sum_{i=0}^{M-2}f_{i}m_{M-i}^{z}\left[  n-1\right]  .
\]
We can use the other memory equations to iterate this procedure, and we end up
with%
\[
\sum_{i=0}^{M}f_{M-i}m_{1}^{z}\left[  n-i\right]  =z\left[  n\right]  .
\]
Noting that%
\[
z^{\prime}\left[  n\right]  =m_{1}\left[  n-M\right]  ,
\]
and using shift-invariance, the above equation becomes%
\begin{equation}
\sum_{i=0}^{M}f_{i}z^{\prime}\left[  n+i\right]  =z\left[  n\right]  .
\label{eq:inf-depth-rec-simple-z}%
\end{equation}
Applying the $D$-transform to (\ref{eq:inf-depth-rec-simple}) gives the
following equation:%
\[
x^{\prime}\left(  D\right)  =\sum_{i=0}^{M}f_{i}D^{i}x\left(  D\right)
=f\left(  D\right)  x\left(  D\right)  .
\]
Applying the $D$-transform to (\ref{eq:inf-depth-rec-simple-z}) gives%
\begin{align*}
\sum_{i=0}^{M}f_{i}D^{-i}z^{\prime}\left(  D\right)   &  =z\left(  D\right) \\
\Rightarrow f\left(  D^{-1}\right)  z^{\prime}\left(  D\right)   &  =z\left(
D\right) \\
\Rightarrow z^{\prime}\left(  D\right)   &  =\frac{1}{f\left(  D^{-1}\right)
}z\left(  D\right)  .
\end{align*}
Rewriting the above set of transformations as a matrix shows that it is
equivalent to the desired transformation in (\ref{eq:z-inf-depth}):%
\[
\left[
\begin{array}
[c]{cc}%
z\left(  D\right)  & x\left(  D\right)
\end{array}
\right]  \left[
\begin{array}
[c]{cc}%
1/f\left(  D^{-1}\right)  & 0\\
0 & f\left(  D\right)
\end{array}
\right]  =\left[
\begin{array}
[c]{cc}%
z^{\prime}\left(  D\right)  & x^{\prime}\left(  D\right)
\end{array}
\right]  .
\]

\end{proof}

Another infinite-depth operation is an \textquotedblleft X\textquotedblright%
\ infinite-depth operation. It transforms the bit representations in
(\ref{eq:logical-operators}) to the following bit representations:%
\begin{equation}
\left[  \left.
\begin{array}
[c]{c}%
0\\
f\left(  D\right)
\end{array}
\right\vert
\begin{array}
[c]{c}%
1/f\left(  D^{-1}\right) \\
0
\end{array}
\right]  , \label{eq:x-inf-depth}%
\end{equation}
where $f\left(  D\right)  $ is defined in (\ref{eq:inf-depth-poly}) and
$1/f\left(  D^{-1}\right)  =D^{M}/D^{M}f\left(  D^{-1}\right)  $.%
%TCIMACRO{\FRAME{ftbpFU}{3.205in}{1.7452in}{0pt}{\Qcb{The quantum shift
%register circuit in the above figure implements the transformation in
%(\ref{eq:x-inf-depth}).}}{\Qlb{fig:x-inf-depth}}{x-inf-depth.pdf}%
%width 3.205in;  height 1.7452in;  depth 0pt;  original-width 6.3529in;
%original-height 3.4333in;  cropleft "0";  croptop "1";  cropright "1";
%cropbottom "0";  filename 'figures/x-inf-depth.pdf';file-properties "XNPEU";}%
%}}%
%BeginExpansion
\begin{figure}
[ptb]
\begin{center}
\includegraphics[
natheight=3.433300in,
natwidth=6.352900in,
height=1.7452in,
width=3.205in
]%
{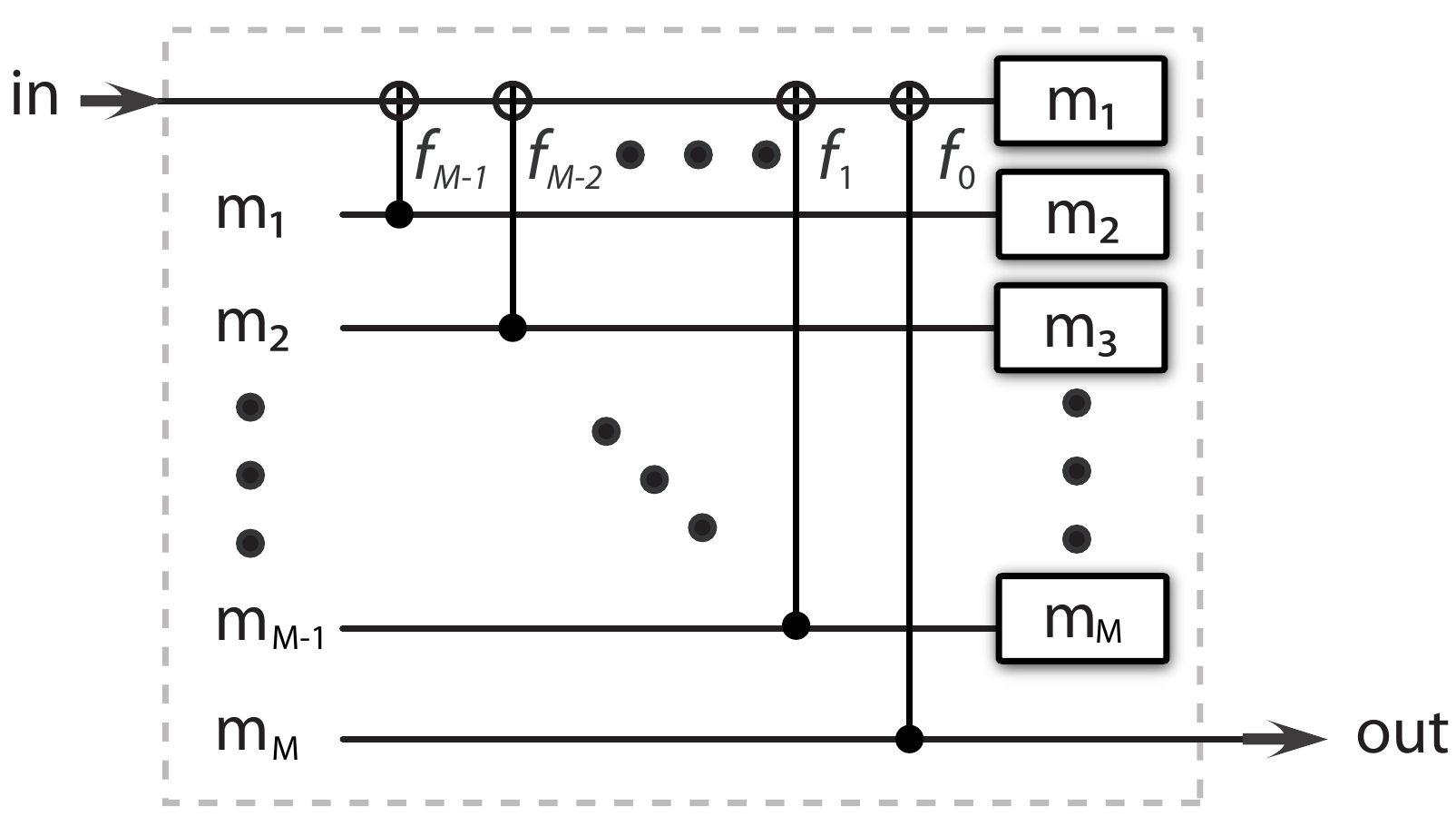}%
\caption{The quantum shift register circuit in the above figure implements the
transformation in (\ref{eq:x-inf-depth}).}%
\label{fig:x-inf-depth}%
\end{center}
\end{figure}
%EndExpansion

\begin{theorem}
The circuit in Figure~\ref{fig:x-inf-depth} implements the transformation in
(\ref{eq:x-inf-depth}) and requires $M$ memory qubits.
\end{theorem}

\begin{proof}
The proof proceeds analogously to the proof of Theorem~\ref{thm:z-inf-depth}%
\ with the \textquotedblleft X\textquotedblright\ and \textquotedblleft
Z\textquotedblright\ variables switching roles because the directionality of
the CNOT gates in the circuit in Figure~\ref{fig:x-inf-depth}\ reverses.
\end{proof}

\section{Memory Requirements for Type II CSS\ Entanglement-Assisted Quantum
Convolutional Codes}

\label{sec:memory-comp-EA}Our last contribution is a formula for the amount of
memory that a Type II CSS\ entanglement-assisted quantum convolutional code. A
Type II CSS\ entanglement-assisted quantum convolutional code is one that uses
infinite-depth operations, outlined in the previous section, in the encoding
circuit \cite{arx2007wildeEAQCC}.

A particular diagonal matrix $\Gamma_{2}\left(  D\right)  $ is the essential
matrix that determines the infinite-depth operations for a Type II
entanglement-assisted code (See Section~VII of Ref.~\cite{arx2007wildeEAQCC}).
Each entry on the diagonal corresponds to an infinite-depth operation, similar
to the polynomial in (\ref{eq:z-inf-depth}) and (\ref{eq:x-inf-depth}).
Therefore, the amount of memory that these infinite-depth operations require
is%
\[
m_{1}\equiv\max_{i}\left\{  \left\vert \deg\right\vert \left(  \left[
\Gamma_{2}\left(  D\right)  \right]  _{ii}\right)  \right\}  .
\]
Suppose a qubit does not have the maximum absolute degree. Alice should delay
this qubit by the difference between that qubit's absolute degree and the
maximum absolute degree so that each qubit lines up properly when output from
the infinite-depth encoding operations. Note that it is not possible to
commute through memory any gates occurring after an infinite-depth operation.

The structure of the encoding circuit for the Type II entanglement-assisted
codes consists of three layers. The first layer is a set of finite-depth
CNOT\ operations characterized by a matrix $L\left(  D\right)  $ (See
Section~VII of Ref.~\cite{arx2007wildeEAQCC}), the second layer consists of
the infinite-depth operations, and the third layer is another set of
finite-depth CNOT\ operations that we name $B\left(  D\right)  $. It is
possible to show that we can implement these encoding circuits with
CNOT\ gates only, as we did in Section~\ref{sec:memory-comp}\ for CSS\ quantum
convolutional codes. Thus, it is straightforward to determine the amount of
memory that a Type II CSS\ entanglement-assisted quantum convolutional code
requires, using Theorem~\ref{thm:memory-CSS}\ and the above upper bound
$m_{1}$

\begin{corollary}
The amount of memory that a Type II CSS\ entanglement-assisted quantum
convolutional code requires is upper bounded by the following quantity:%
\[
m_{1}+\left\vert \deg\right\vert \left(  L\left(  D\right)  \right)
+\left\vert \deg\right\vert \left(  B\left(  D\right)  \right)  .
\]

\end{corollary}

\section{Conclusion}

I have developed the theory for a quantum shift register circuit. These
circuits can encode and decode quantum convolutional codes. The two important
contributions of this paper are the technique for cascading quantum shift
register circuits and the formulas for the memory required by a CSS\ quantum
convolutional code. Quantum shift register circuits should be useful for
experimentalists wishing to demonstrate the operation of a quantum
convolutional code.

Some interesting open questions remain. I have not yet determined the amount
of memory that a general (non-CSS) code requires. The proof technique of
Theorem~\ref{thm:memory-CSS}\ does not extend to combinations of
controlled-Phase and controlled-NOT gates because they combine differently
from the way that cascades of controlled-NOT\ gates combine. It might also be
interesting to study the entanglement structure of states that are input to a
quantum shift register circuit, in a way similar to the observation in
Ref.~\cite{arx2008wildeOEA}\ concerning the relation of an entanglement
measure to an entanglement-assisted code.

\begin{acknowledgments}
I\ thank Martin R\"{o}tteler for the initial suggestion to pursue a quantum
shift register implementation of encoding circuits for quantum convolutional
codes, for hosting me as a visitor to NEC\ Laboratories America for the month
of September 2008, and for useful comments on the presentation of this
manuscript. I thank Martin R\"{o}tteler, Hari Krovi, and Markus Grassl for
useful discussions on this topic. I thank Kevin Obenland and Andrew Cross for
discussions on this topic and thank Kevin Obenland for the suggestion to make
the quantum shift register diagrams \textquotedblleft read like a
book,\textquotedblright\ from left to right and top to bottom. I acknowledge
support from the internal research and development grant SAIC-1669 of Science
Applications International Corporation.
\end{acknowledgments}

\bibliographystyle{unsrt}
\bibliography{Ref}

\end{document}